\documentclass[sn-nature]{sn-jnl}

\usepackage{amsmath,amssymb,amsfonts}
\usepackage{amsthm}
\usepackage{mathrsfs}

\usepackage{graphicx}
\usepackage{booktabs}
\usepackage{tabularx}
\usepackage{multirow}
\usepackage[labelfont=bf,textfont=normalfont]{caption}
\usepackage{subcaption}

\usepackage{siunitx}
\usepackage{gensymb}
\usepackage{textcomp}

\usepackage[utf8]{inputenc}
\usepackage{newunicodechar}
\newunicodechar{−}{-}

\usepackage{algorithm}
\usepackage{algorithmicx}
\usepackage{algpseudocode}
\usepackage{listings}

\usepackage[title]{appendix}
\usepackage{xcolor}
\usepackage{manyfoot}
\usepackage{setspace}

\usepackage{bibunits}
\defaultbibliographystyle{sn-nature}
\defaultbibliography{references}

\usepackage{hyperref}

\theoremstyle{thmstyleone}

\theoremstyle{thmstyletwo}

\theoremstyle{thmstylethree}

\unnumbered
\begin{document}
\begin{bibunit}[sn-nature]

\title{A neural network-based Universal Thermal Climate Index for reliable global thermal-stress classification across extreme weather}


\author*[1]{\fnm{Bikem} \sur{Pastine}}\email{bikem.pastine@ouce.ox.ac.uk}
\author[2]{\fnm{Milan} \sur{Klöwer}}
\author[3]{\fnm{Tianning} \sur{Tang}}
\author[1]{\fnm{Sarah} \sur{Wilson Kemsley}}
\author[1]{\fnm{Louise} \sur{Slater}}

\affil*[1]{{\small \orgdiv{School of Geography and the Environment, University of Oxford, UK}}}

\affil[2]{{\small \orgdiv{Atmospheric, Oceanic and Planetary Physics, University of Oxford, UK}}}

\affil[3]{{\small \orgdiv{Department of Mechanical and Aerospace Engineering, University of Manchester, UK}}}


\abstract{
Extreme temperatures are the leading cause of climate-related mortality worldwide. Climate-health research and operational weather forecasting require accurate estimates of human thermal stress. The Universal Thermal Climate Index (UTCI) is among the most sophisticated and widely used feels-like temperature metrics. However, its ubiquitous polynomial approximation does not generalize well to extreme weather conditions. Here, we introduce Neural-UTCI, a neural network that calculates UTCI with substantially higher accuracy across global conditions at a lower computational cost for operational use. Neural-UTCI reduces the polynomial approximation RMSE from 2.78 °C to 0.36 °C, an 87\% improvement, and lowers thermal stress misclassification rates from 5.3\% to 1.7\%, with consistent performance across resampling experiments. These differences affect thermal exposure metrics. For example, during the 2003 European heatwave summer in Rome, Italy, the number of very strong heat stress days increases from 15 to 35 days when using Neural-UTCI compared to operational products like ERA5-HEAT. Simultaneously, Neural-UTCI reliably classifies extreme cold stress conditions, allowing continuous global application. By improving UTCI accuracy, Neural-UTCI can strengthen climate-health risk assessments and public weather warning systems, especially as global warming increases the incidence of extreme events.
}


\maketitle
\section{Introduction}

Non-optimal temperatures present a major health risk, causing about 5 million excess deaths per year globally \cite{Zhao2021} but the mortality burden of both cold and heat events can be mitigated if they are anticipated \cite{Kotharkar2022}. \textit{Feels-like temperature} metrics are used to predict human outcomes in response to weather conditions \cite{Freitas2017}. Indices like wet bulb temperature and wind chill use simple functions to summarize the joint effect of key physiologically-relevant environmental variables. Though computationally inexpensive, they only crudely represent physiological mechanisms. The Universal Thermal Climate Index (UTCI) is a process-based model that represents human heat transfer and thermoregulation to map global environmental conditions to equivalent thermal stress responses \cite{Jendritzky2012}. UTCI is extensively validated and is among the most robust feels-like temperature indices \cite{Havenith2024}, featuring in over 1,000 journal articles.  

UTCI predicts equivalent thermal response using the Fiala multi-node model coupled to a clothing model (which together we refer to as Fiala-UTCI) with four meteorological inputs: air temperature, humidity, wind speed, and mean radiant temperature \cite{Brode2012}. The Fiala model is computationally intensive \cite{DiNapoli2021} and is proprietary paid software. In practice, UTCI is therefore exclusively calculated using a sixth-order polynomial approximation (hereafter Polynomial-UTCI) developed by Bröde et al. (2012) \cite{Brode2012}.  Common implementation tools such as RayMan \cite{Meng2025}, BioKlima 2.6 \cite{Blazejczyk2017BioKlima}, pythermalcomfort \cite{Tartarini2020}, thermofeel \cite{Brimicombe2022thermofeel}, and the official UTCI FORTRAN calculator \cite{Brode2009utciwebsite} all rely internally on Polynomial-UTCI and not Fiala-UTCI. Polynomial-UTCI is used across many domains: from health studies \cite{Romaszko2022} to outdoor thermal planning \cite{Zhao2025}. Most notably, Polynomial-UTCI is used for forecasting population-level thermal health hazards \cite{Romaszko2022} by five national weather centers \cite{DiNapoli2021} and is forecast globally by the European Centre for Medium-Range Weather Forecasts (ECMWF) \cite{DiNapoli2020}. Additionally, the ECMWF hosts ERA5-HEAT which is a long-term global hourly Polynomial-UTCI product with a 0.25° resolution. ERA5-HEAT is used to study trends in heat stress exposure \cite{Emerton2026,Nasara2025, Hamed2025, Grigorieva2023} and the associated health, economic, and political impacts under climate change \cite{Shah2025,Li2025,Hoffmann2022}. 

Because of the ubiquitous use of Polynomial-UTCI, the scientific and forecasting literature conflates the approximation with the full original model though the distinction between Fiala-UTCI and its approximation can be consequential. For instance, when wind speed is above 17 m/s the index is considered out-of-domain because Polynomial-UTCI is known to poorly approximate Fiala-UTCI in this range  (RMSE 9.0°C) \cite{Brode2012, Pappenberger2015}. The error associated with Polynomial-UTCI in the remaining input range is rarely acknowledged: the skill of UTCI forecasts is assumed to be determined by the skill of the forecasting system \cite{Pappenberger2015} despite the non-negligible error from the approximating polynomial (RMSE 1.1°C). Some environmental and health studies find that UTCI poorly predicts human thermal response, even at moderate wind speeds well below the 17 m/s threshold and cool temperatures \cite{Staiger2019, Nie2022, Urban2014} but the literature attributes this limitation to a bias in the underlying Fiala-UTCI model which has no reported wind bias \cite{Psikuta2012}. An improved operational UTCI approximation with higher fidelity to Fiala-UTCI would allow the index to be applied in extreme conditions where biases in Polynomial-UTCI may be affecting scientific results without the knowledge of its users.  

Since Polynomial-UTCI was developed in 2012 \cite{Brode2012}, new modeling methods have become standard tools in environmental and health sciences. UTCI has recently been re-derived from thermal response data using modern statistical methods by Bröde et al. (2024) to affirm the index design but not to propose a more accurate operational model \cite{Brode2024}.  UTCI downscaling, postprocessing, and local prediction models have been developed using machine learning tools, but these studies have maintained use of Polynomial-UTCI for their initial calculation \cite{Alinasab2025, Yang2024, Kuzmanovic2024, Gong2024, Briegel2024, Liu2022}.  Machine learning is used to improve local UTCI heat early-warning systems as part of data-driven weather models but the models are trained on ERA5-HEAT \cite{Collazo2026}. Roman et al. (2026) \cite{Roman2026} show that sparse regression using an orthogonal Legendre polynomial basis yields a more numerically stable and robust approximation of Fiala-UTCI than the original sixth-order polynomial. However, accuracy improvements are modest because the paper aims to demonstrate the benefits of sparse regression for preserving interpretability and computational speed rather than to improve the approximation for operational use. 

In this study we prioritize UTCI approximation accuracy for operational use. We develop a compact neural-network approximation (Neural-UTCI) to Fiala-UTCI as a drop-in replacement for Polynomial-UTCI with increased accuracy that is widely applicable across weather conditions and operational procedures. Our design has similar computational requirements on CPUs and greatly increased performance on GPUs for modern use and is published as an open-source Python package. The accuracy improvements of Neural-UTCI are related to a better reflection of physiological processes. We find that these improvements yield substantial downstream impacts. During Europe’s most deadly extreme temperature events, for instance, differences between Neural-UTCI and Polynomial-UTCI thermal stress categorization highlight the added uncertainty introduced by the current operational procedures when issuing public heat warnings. We further find that Polynomial-UTCI  introduces systematic and spatially structured errors in global 2024 UTCI estimates compared to Neural-UTCI, altering global thermal stress exposure estimates.

\section{Data and Methods}
\subsection{Datasets}\label{datasets}
In order to build approximation models of Fiala-UTCI, we use the same Fiala-UTCI simulations published with the Polynomial-UTCI operational procedure \cite{Brode2012} (see Supplementary Methods 1). The Fiala-UTCI model is the intellectual property of Ergonsim, a human thermal modeling company, and the full model is not freely available for public use. Generating new Fiala-UTCI data is therefore not feasible and these are the only available reference datasets for developing new approximations. 

We use the 'Grid data' from the ESM4 of Bröde et al.\cite{Brode2012} to train the models. The dataset comprises 104,643 Fiala-UTCI simulations sampled incrementally from the function's input domain.  We also use the independent ECHAM4 testing dataset published by Bröde et al.\ (2012 and 2024) \cite{Brode2012,Brode2024} to evaluate UTCI approximations. This is an independent sample of Fiala-UTCI results for 1,000 randomly selected meteorological conditions from a control run (1971–1980) of the European Centre Hamburg Model version 4 (ECHAM4) general circulation model \cite{Stendel1998}. We downloaded the ECHAM4 dataset from Zenodo \cite{Brode2021}. Details on the input domain for both datasets are described in Supplementary Methods S2 and the distributions are visualized in Supplementary Figure S1. 

We compared the patterns in prediction differences between Polynomial-UTCI and our best performing approximating function, Neural-UTCI, using ECMWF ERA5 reanalysis data \cite{Hersbach2020ERA5}.  We used hourly single level ERA5 data to calculate Neural-UTCI at a spatial resolution of 0.25° by 0.25° and compared the results to ERA5-HEAT UTCI, which evaluates Polynomial-UTCI on the same ERA5 reanalysis input \cite{DiNapoli2020}. We use mean radiant temperature produced for ERA5-HEAT as an input for Neural-UTCI, the calculation of which is detailed in Di Napoli, 2020 \cite{DiNapoli2020MRT}. We downloaded all ERA5 data from the Copernicus Climate Change Service. Variable names and preprocessing steps are described in Supplementary Methods S3. 

\subsection{Model Development}
Following Bröde et al.\ (2012 and 2024), we trained and validated on the grid dataset and held out the ECHAM4 data for independent external testing.  We applied a 90/10 random split on the grid data for training ($n=94,178$) and validation ($n=10,465$). We chose a 90/10 split to maximize the use of available data though other random split proportions decreased model performance modestly. 

We compared candidate regression methods to select the most appropriate approximating method. We compared polynomial multiple linear regression and ridge regularized regression (2nd to 8th order, inclusive), support vector regression, random forest regression, XGBoost, and a feed-forward neural network. Optimal forms of the models were found using Bayesian search. Models were compared using several verification metrics, detailed in Supplementary Methods S4. 

We found a compact feed-forward neural network to be the best performing method and the model was further refined using hyperparameter optimization (Supplementary Methods S5). We found the close-optimal neural network architecture to be (4 inputs, 79, 75, 39, 1 output) with ReLU activation functions and mean squared error as the loss function. The hyperparameters selected after tuning can be found in Supplementary Table S7. We release the optimized approximation model, Neural-UTCI as a python package \cite{Pastine2026}. 

\subsection{Quantifying model error}
We estimated the interquartile ranges (IQR) for RMSE and miscategorisation rates by drawing 1,000 bootstrap samples of the testing data, each consisting of 500 observations sampled with replacement to provide a measure of how sensitive these metrics were to the composition of the testing dataset. 

We use the variance sum law to approximate the contribution of the UTCI approximation error to the total UTCI error, assuming unbiased, independent, and normally distributed errors. We let the error associated with Fiala-UTCI equal the reported RMSE for dynamic thermal sensation averaged over simulated exposure times: 1.04°C \cite{Broede2012Brazil}. Further details of assumptions and methodology can be found in Supplementary Methods S6.

\subsection{Comparing approximator computational speed} \label{comp_speed} 
We compared the computational speed of the two UTCI approximation methods without preprocessing steps. We measured the execution time required to evaluate one million individual and batched input samples on a CPU for both methods and on a GPU for Neural-UTCI.  We compared both CPU and GPU performance of Neural-UTCI to only the CPU performance of Polynomial-UTCI because only Neural-UTCI offers this advantage natively. Results are shown in Supplementary Table S3.

\subsection{SHAP analysis} \label{shap}
We conducted SHapley Additive ExPlanation (SHAP)  \cite{Lundberg2017} analysis to compare model behavior and feature importance between approximators. We used the \textit{shap} package in Python and the shap.Explainer function. We preferred the \textit{explainer} method to comparably calculate SHAP between models, both initialized with the training subset of the grid data, and SHAP values are evaluated for the ECHAM4 testing data.

\subsection{Heat and cold stress exposure metrics} \label{heat_exposure}
We calculate Heat Stress Exposure as the number of hours at each ERA5 grid cell where UTCI exceeds the strong heat stress category (\(>\)32°C) as described in Shah et al. (2025) \cite{Shah2025} and Cold Stress Exposure as the number of hours under the threshold for strong cold stress (\(<\)-13°C). 

\subsection{European extreme event case study selection} \label{case_study}
We assessed potential operational differences between Polynomial-UTCI and Neural-UTCI by examining two of the most lethal extreme temperature events in Europe: the 2003 heatwave (70,000 deaths \cite{Robine2008}) and the 2012 cold spell (800 deaths \cite{Bowen2012}).  For the 2003 heatwave, we calculated approximator differences for June through August 2003 following the period set by Stott et al. (2004) \cite{Stott2004}, in order to capture geographic variation in UTCI differences during the local peak of the heatwave in each location. 

The dates for the 2012 European cold spell are location and identification method dependent, with the most conservative estimates identifying a 10-day period between 3 and 12 February \cite{Ribes2025} and the most liberal estimates identifying a 21-day period starting in late January \cite{Lan2013}. To capture as much geographic variation in approximator differences as possible, we define the cold period as January through February 2012.

\section{Results and Discussion}

\subsection{A neural network-based Universal Thermal Climate Index} 
\begin{figure}[H] 
    \centering
    \includegraphics[width=1\textwidth]{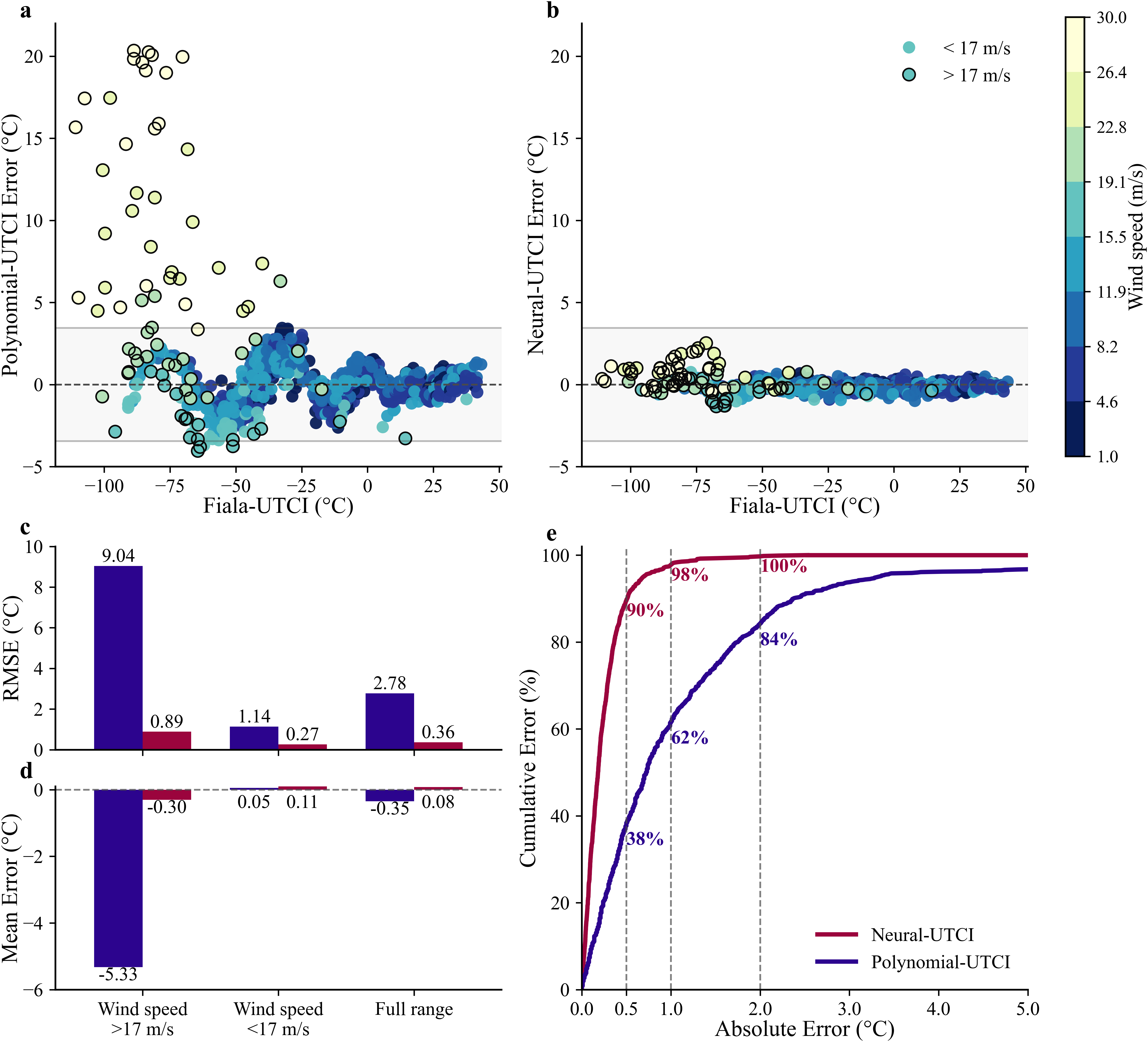}
    \caption{\textbf{Performance of the two approximating models on the Fiala-UTCI testing dataset.} \textbf{(a-b)} Polynomial and neural network model approximation error relative to Fiala-UTCI (approximator minus Fiala-UTCI, markers with black edges denote wind speed above 17 m/s). Model root mean squared error \textbf{(c)} and mean error (as defined in Supplementary Methods S1) \textbf{(d) } comparison between Polynomial-UTCI and Neural-UTCI where wind speed is $ < $17 m/s, $>$17 m/s, and the full wind speed range. \textbf{(e)} Cumulative error distributions of the approximators. Vertical dashed lines and labels denote the percentage of the dataset with absolute errors under 0.5, 1.0, and 2.0~\textdegree{}C.} 
    \label{utci_perf}
\end{figure}

Neural-UTCI substantially reduces emulation error relative to Polynomial-UTCI across the range of Fiala-UTCI data (Fig. \ref{utci_perf}a and \ref{utci_perf}b), with an RMSE of  0.36°C (IQR 0.35–0.38°C) compared to 2.78°C (IQR 2.53–2.99°C). Approximation error is the dominant contributor to total UTCI error for Polynomial-UTCI, but not for Neural-UTCI (Supplementary Figure S2). Relative to the Fiala-UTCI model which has a reported 1.04°C RMSE \cite{Broede2012Brazil}, Neural-UTCI's 0.36°C approximation error contributes only 19\% to total operational UTCI error, whereas Polynomial-UTCI increases expected total error by 132\%. 

Neural-UTCI reduces the occurrence of large approximation errors, especially at high wind speeds. All Neural-UTCI errors lie within ±2°C in the testing data, whereas 16\% of cases exceed this range using Polynomial-UTCI (Fig. \ref{utci_perf}e). Moreover, 90\% of Neural-UTCI errors fall within ±0.5°C in contrast to only 38\% for Polynomial-UTCI (Fig. \ref{utci_perf}e). Under high-wind conditions  ($>$17 m/s), Polynomial-UTCI performance deteriorates substantially (RMSE = 9.03°C, IQR 8.82-9.26°C), with individual UTCI errors exceeding 20°C (Fig. \ref{utci_perf}a). Neural-UTCI reduces the high wind RMSE to 0.89°C (IQR 0.86-0.91°C) (Fig. \ref{utci_perf}c), enabling more reliable calculations across the full range of Fiala-UTCI testing data. 

The two UTCI approximation methods have a negligible speed difference for individual and batched inputs on the CPU, but Neural-UTCI can compute batched inputs 33 times faster on a GPU than the CPU-based polynomial implementation (Supplementary Table S3), depending on batch size and GPU specifications. For operational uses, large batched UTCI calculations can be made substantially faster using Neural-UTCI on a GPU than was possible with Polynomial-UTCI.


\subsection{Improved accuracy of thermal stress categorisation} 

UTCI can be classified into thermal stress categories using physiological thresholds (Supplementary Methods S1) \cite{Jendritzky2012}. These categories support weather warnings and public health guidance \cite{Blazejczyk2010} and are used in climate research to assess exposure to thermal stress \cite{Shah2025, Peng2026}. Because categorisation is threshold-based, UTCI accuracy is especially important near category boundaries, where small approximation errors can change the assigned category. We therefore evaluate the category-level performance of Neural-UTCI and Polynomial-UTCI relative to Fiala-UTCI.

\begin{figure}[H] 
    \centering
    \includegraphics[width=1.1\textwidth]{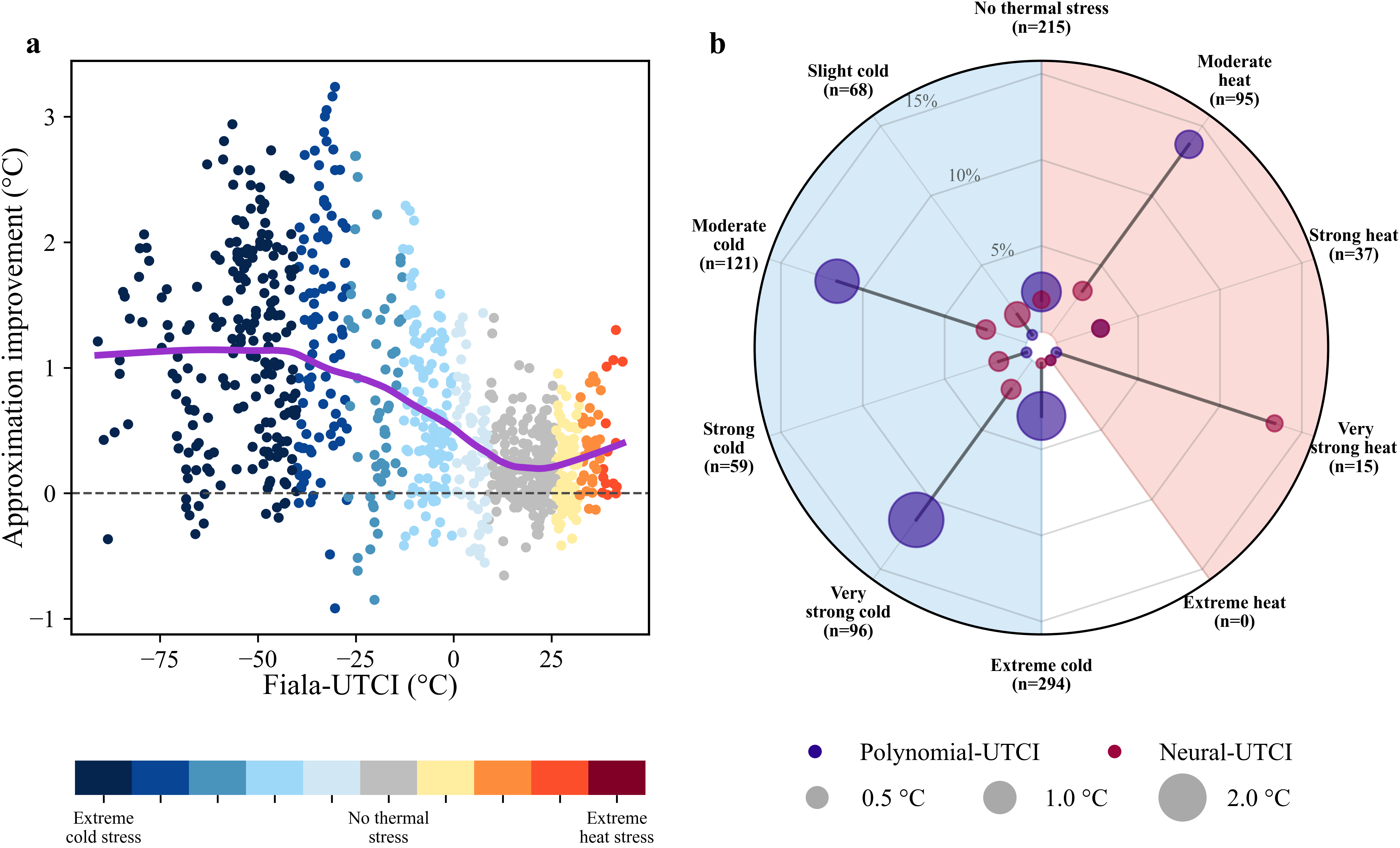}
    \caption{\textbf{UTCI approximation error by thermal stress category.} \textbf{(a)}  Approximation improvement of using the neural network compared to the polynomial at wind speeds under 17 m/s. Approximation improvement is defined as the reduction of absolute error. The purple line denotes a LOESS fit (span = 0.4). Points are colored by Fiala-UTCI thermal stress category (Supplementary Methods S1). \textbf{(b)} Thermal stress misclassification rate using the neural network and polynomial approximators. Light red shaded sectors denote heat stress and light blue sectors denote cold stress categories. Sector labels include the number of observations (n) in each thermal stress category. The distance of points from the center represents the misclassification rate for the category. The sizes of the points represent the average absolute approximation error (°C) of the misclassified instances. Black lines connecting circles highlight the difference in misclassification error between Polynomial-UTCI and Neural-UTCI.} 
    \label{miscat}
\end{figure}

Neural-UTCI has lower absolute error than Polynomial-UTCI in 85\% of test cases, with lower errors across all thermal stress categories, reducing approximation error by more than 1°C in 35\% of cases.  (Fig. \ref{miscat}a, Supplementary Figure S3). The largest improvement occurs in the \textit{extreme cold stress} category, with a mean reduction in approximation error of 2.4°C, whereas the smallest improvement is observed in the \textit{no thermal stress} category, with a mean reduction of 0.2°C (Fig. \ref{miscat}a). Among the heat stress categories, \textit{very strong heat stress} shows the largest benefit from Neural-UTCI. Although the mean approximation improvement is only 0.4°C in this category, small UTCI errors may affect downstream applications such as heat-related mortality risk estimation \cite{Kuchcik2021}, particularly where risk increases exponentially with heat exposure \cite{Gasparrini2011}. 

Categories assigned using Neural-UTCI agree more closely with the Fiala-UTCI classification than those assigned using Polynomial-UTCI (Fig. \ref{miscat}b). The overall misclassification rate is 1.7\% (IQR 1.4-2.0\%) for Neural-UTCI and 5.3\% (IQR 4.8-5.7\%) for Polynomial-UTCI, representing a 68\% reduction in misclassification frequency. Of the cases misclassified by Neural-UTCI, 60\% are also misclassified by Polynomial-UTCI. Polynomial-UTCI performs better in heat stress than in cold stress conditions, but the testing dataset contains few heat stress observations; specifically, there are no \textit{extreme heat stress} observations and only a limited sample (\textit{n}=15) of \textit{very strong heat stress} cases. In the training data, Polynomial-UTCI misclassifies 21\% of \textit{extreme heat stress} observations and 14\% of \textit{very strong heat stress} observations, compared with 0.6\% and 2\%, respectively, for Neural-UTCI (Supplementary Figure S3). Neural-UTCI misclassifies at a low and broadly uniform rate across categories, whereas Polynomial-UTCI has a less reliable performance and larger UTCI errors when misclassification occurs (Fig. \ref{miscat}b).

\subsection{More physiologically consistent representation of humidity and wind effects}
Using SHAP (Methods \ref{shap}), we evaluate whether the functional behavior of Neural-UTCI and Polynomial-UTCI is consistent with the physiological relationships described by the Fiala model. Similar decompositions have been used to predict thermal-stress health risks at sporting events \cite{Klower2023, Hollander2021} but, to our knowledge, no prior work has examined the functional behavior of Polynomial-UTCI as an approximator by input parameter. It is important that the relationships captured by the approximators are physiologically plausible for the model to be trustworthy, beyond single-number performance metrics.

\begin{figure}[H] 
    \centering
    \includegraphics[width=1\textwidth]{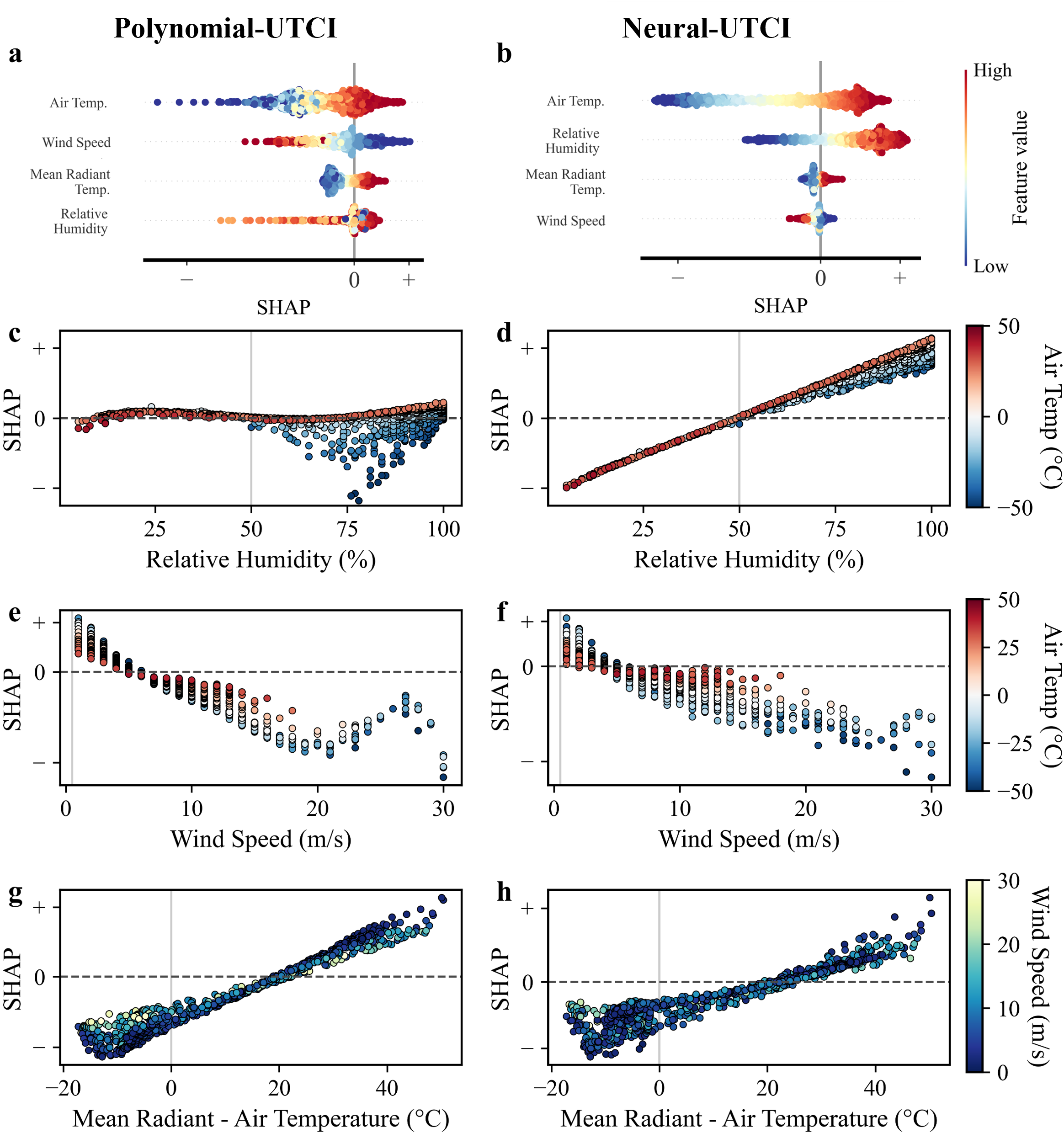}
    \caption{\textbf{Learned predictor-response relationships in the polynomial and neural network UTCI approximator models using SHapley Additive ExPlanation (SHAP) analysis.} \textbf{(a-b)} Global feature importance and the direction of effect. Importance of each predictor is ranked, the horizontal position denotes the marginal contribution. The relationship between UTCI offset (°C) and the input variables air temperature (°C), wind speed (m/s), Mean Radiant Temperature (°C), and relative humidity (\%) are shown. Positive SHAP values imply a positive effect of increasing the variable on the prediction and vice versa. Only the sign of SHAP values is shown (magnitudes not comparable between models). \textbf{(c-d)} SHAP dependence plots showing the effect of relative humidity on the UTCI output. \textbf{(e-f)} As (c-d), but for wind speed. \textbf{(g-h)} As (c-d) but for mean radiant temperature minus air temperature. Gray vertical lines denote the reference condition as defined by Fiala-UTCI.} 
    \label{3}
\end{figure}

Both Neural-UTCI and Polynomial-UTCI  identify air temperature as the most important predictor of UTCI offset, but differ on the importance assigned to the remaining predictors (Fig.~\ref{3}a,b). 

Relative humidity is an important determinant of heat stress because high humidity reduces evaporative cooling from sweating \cite{Dzyuban2020}. Relative humidity has the lowest predictor importance in Polynomial-UTCI (Fig.~\ref{3}a). At high temperatures, Polynomial-UTCI varies little with relative humidity, whereas at low temperatures, high relative humidity reduces Polynomial-UTCI (Fig.~\ref{3}c). In contrast, Neural-UTCI identifies relative humidity as its second most important predictor after air temperature (Fig.~\ref{3}b). Relative humidity below the reference condition of 50\% reduces Neural-UTCI, while humidity above 50\% increases it (Fig.~\ref{3}b). Neural-UTCI therefore captures the relationship between humidity and thermal stress represented by the Fiala model \cite{Fiala2012} while Polynomial-UTCI does not.

At low temperatures, higher wind speeds increase sensible heat loss from the body, producing the wind-chill effect \cite{Zhao2024}. As air temperature approaches skin temperature, sensible heat loss becomes less efficient, although higher wind speeds increase evaporation and thereby enhance latent heat loss \cite{osczevski1995windchill}. Both Polynomial-UTCI and Neural-UTCI  associate higher wind speeds with lower UTCI values (Fig.~\ref{3}a,b), with the magnitude of this effect depending on air temperature (Fig.~\ref{3}e,f). The air temperature dependence is more pronounced in Neural-UTCI than in Polynomial-UTCI, with only a marginal contribution of wind speed at high air temperatures but a substantially stronger effect at low temperatures (Fig.~\ref{3}f). Polynomial-UTCI diverges from the expected relationship at wind speeds of 20~m\,s$^{-1}$, likely contributing to its large errors under high-wind conditions  (Fig.~\ref{3}e).

\subsection{Approximator choice alters estimated thermal stress during extreme events}
We assess differences between Polynomial-UTCI and Neural-UTCI during the 2003 European heatwave (June–August 2003) and 2012 European cold spell (January–February 2012) (Methods \ref{case_study}).  

\begin{figure}[H] 
    \centering
    \includegraphics[width=1\textwidth]{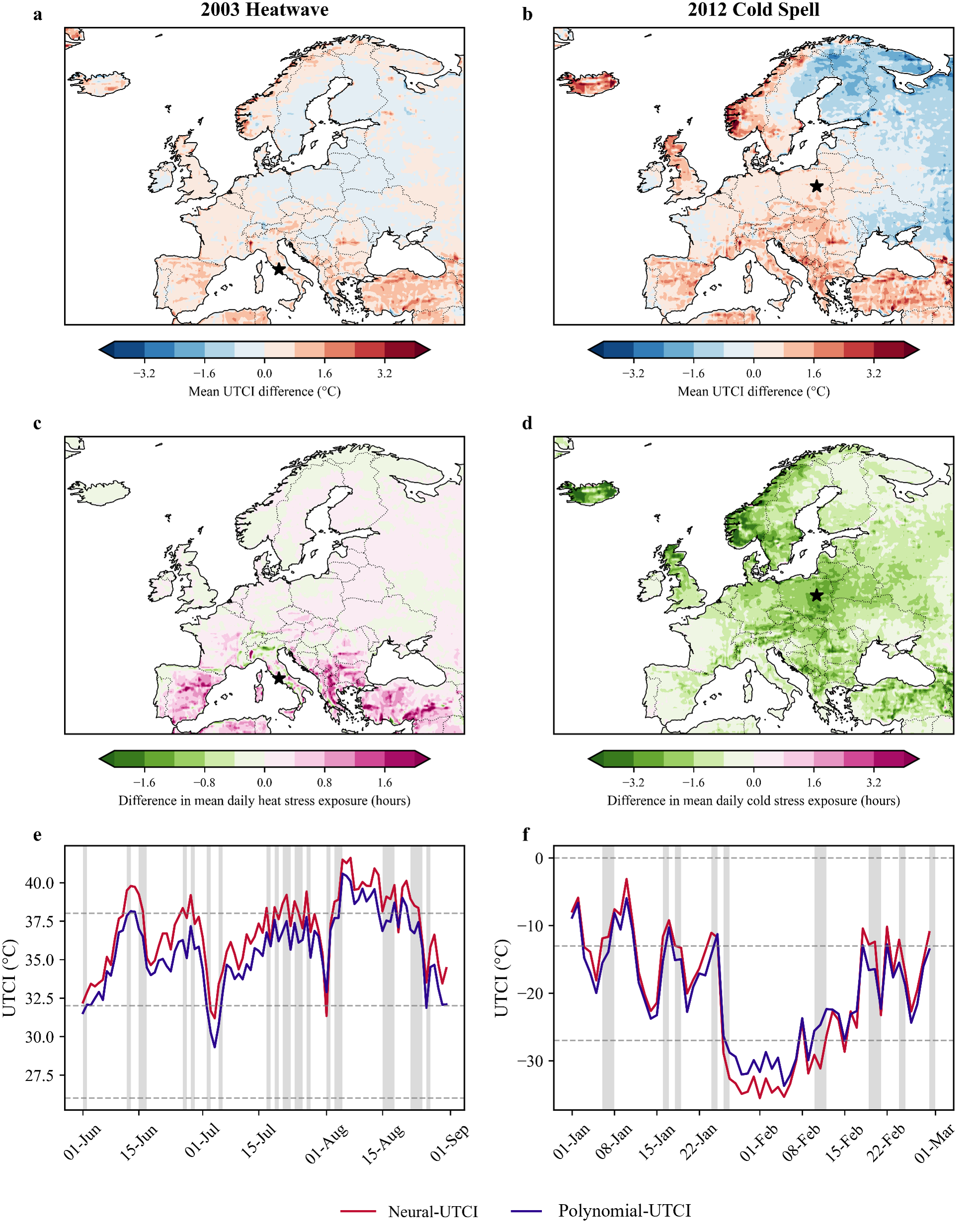}
    
    \caption{\textbf{Difference in UTCI predicted by Neural-UTCI and Polynomial-UTCI for the 2003 European heatwave and the 2012 European cold spell.} \textbf{(a-b)} Mean difference in hourly UTCI per grid cell between the two approximations (Neural-UTCI and Polynomial-UTCI) for 1 June to 1 September 2003 and 1 January to 1 March 2012, respectively. Stars mark Rome, Italy and Warsaw, Poland. \textbf{(c-d)} Average difference in mean daily thermal stress exposure times over the same period, calculated as daytime hours with UTCI $>$32°C (heat stress) and $<$-13°C (cold stress). \textbf{(e-f)} Daily maximum \textbf{(e)} and minimum\textbf{ (f)} UTCI in Rome and Warsaw, respectively, from Neural-UTCI (red) and Polynomial-UTCI (blue). Horizontal dashed lines show UTCI category thresholds for 'moderate', 'strong', and 'very strong' thermal stress categories of heat stress (ascending order, panel \textbf{(e)}), and cold stress (descending order \textbf{(f)}).  Gray shading marks days where the thermal stress categories calculated by the approximating functions are different.}
    \label{4}
\end{figure}

During the 2003 heatwave, Neural-UTCI estimates mean UTCI  0.16°C higher than Polynomial-UTCI, but this spatial mean masks substantial regional differences. The largest approximator differences are observed in the Mediterranean, where Neural-UTCI estimates mean UTCI 0.40°C higher than Polynomial-UTCI (Fig.~\ref{4}a). These small absolute UTCI differences nevertheless affect calculated thermal stress exposure (Fig.~\ref{4}c). Neural-UTCI estimates approximately 0.36 additional hours of heat-stress exposure per day in the Mediterranean relative to Polynomial-UTCI, totaling a grid cell average of 34 additional heat-stress hours over the summer  (Fig.~\ref{4}c). 

During the 2012 cold spell, the mean difference across Europe is also small, with Neural-UTCI estimating mean UTCI 0.02°C higher than Polynomial-UTCI. However, the direction and magnitude of the differences vary spatially. In the Mediterranean region, Neural-UTCI estimates temperatures that are 0.72°C warmer than Polynomial-UTCI, whereas in north-eastern Europe, which was especially affected by the 2012 cold spell \cite{Demirtas2017}, Neural-UTCI estimates mean UTCI 0.76°C lower than Polynomial-UTCI.  Across Europe, Neural-UTCI estimates 1.26 fewer hours of cold-stress exposure per day, equivalent to a mean of 76 fewer hours per grid cell over the two-month period (Fig. \ref{4}d). 

Together, these events show that approximator choice can affect estimated thermal-stress exposure despite small mean UTCI differences, with effects that vary substantially across regions. 

To assess potential implications for UTCI-based thermal-stress warnings, we examine daily UTCI maxima and minima in Rome, Italy, during summer 2003 and Warsaw, Poland, during the 2012 cold spell. UTCI is used in feels-like temperature forecasting in both countries \cite{Maracchi2017,Blazejczyk2010}. Neural-UTCI estimates higher mean UTCI than Polynomial-UTCI for both case studies: 1.3°C higher in Rome during the heatwave and 0.2°C higher in Warsaw during the cold spell (Fig.~\ref{4}e,f). When thermal-stress categories are calculated, the approximators assign 27\% of days to different categories during summer 2003 in Rome. Neural-UTCI increases the number of \textit{very strong heat stress} days from 15 to 35. During the 2012 cold spell in Warsaw, the approximators assign 20\% of days to different categories. Neural-UTCI classifies three additional days as \textit{very strong cold stress}, while reclassifying nine days from \textit{strong cold stress} under Polynomial-UTCI to \textit{moderate cold stress}. These discrepancies are operationally relevant because public-health guidance is category-dependent. For example, Polish guidance recommends limiting outdoor exposure under \textit{strong cold stress}, whereas under \textit{very strong cold stress} it issues a frostbite warning  \cite{Blazejczyk2010}.

\subsection{Systematic global differences between UTCI approximators } 

We illustrate geographic differences between Neural-UTCI and Polynomial-UTCI at a global scale by evaluating both models on hourly ERA5 data from 2024. 

Unlike Polynomial-UTCI, Neural-UTCI can estimate UTCI across the full wind-speed range represented in ERA5. ERA5-HEAT therefore produces incomplete UTCI time series in 43\% of global land and ocean grid cells, with a mean of 132 missing hours per affected grid cell in 2024 which can be corrected by Neural-UTCI.

\begin{figure}[H] 
    \centering
    \includegraphics[width=0.9\textwidth]{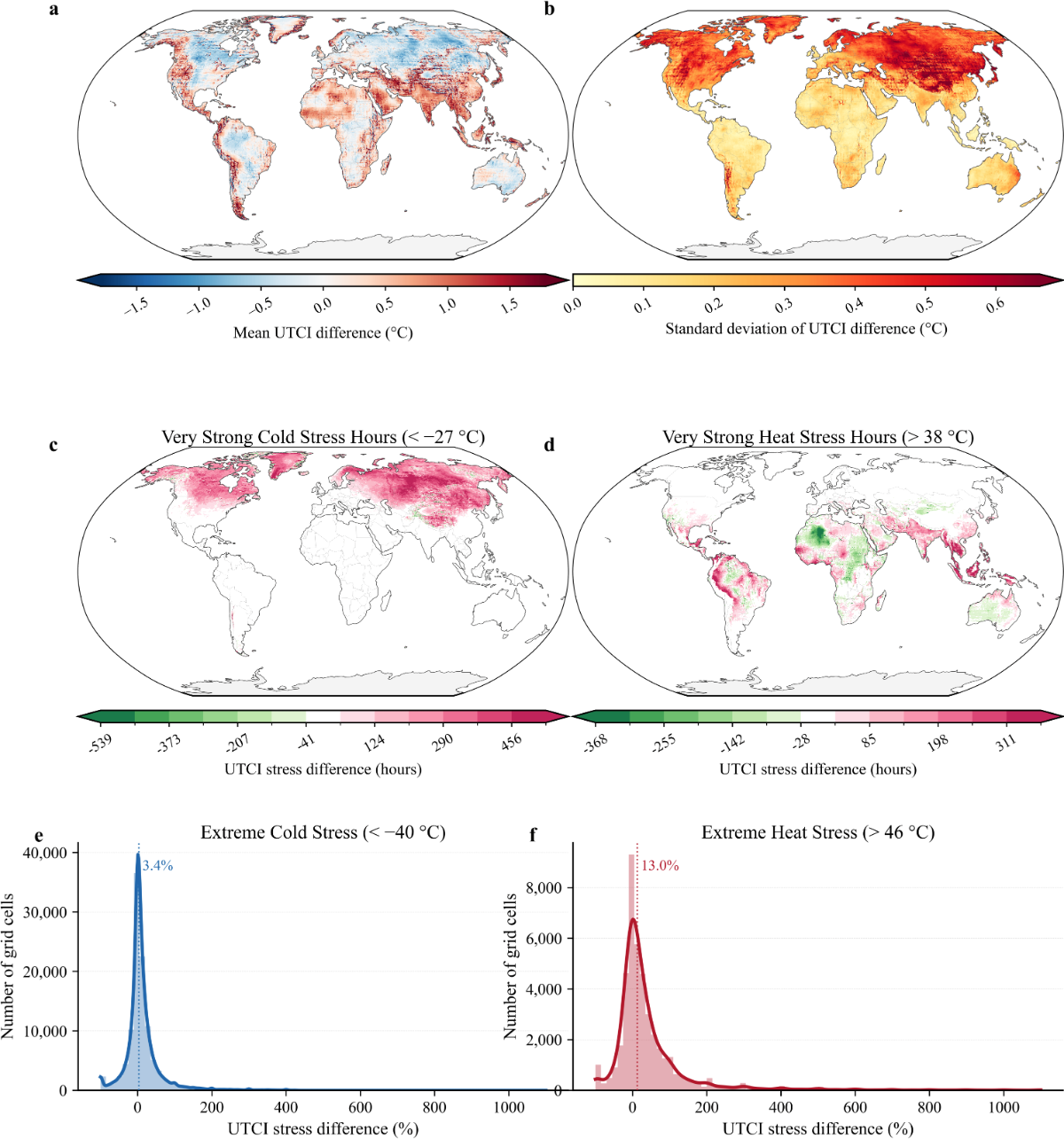}
    
    \caption{\textbf{Difference in UTCI predicted by Neural-UTCI and Polynomial-UTCI in 2024.} The mean \textbf{(a)} and standard deviation \textbf{(b)} of difference between the UTCI approximating functions using hourly ERA5 data for 2024. Difference in the number of hours spent below the very strong cold stress threshold \textbf{(c)} and above very strong heat stress threshold \textbf{(d)} at each grid cell. Density plots showing the distribution of percentage difference of hours at extreme cold \textbf{(e)} and extreme heat \textbf{(f)}. The histogram density represents the number of grid cells with a given percentage stress-hour difference. The dotted vertical line shows the median difference.}
    \label{fig:5_2024}
\end{figure}

Approximator differences vary geographically and with temperature (Fig. \ref{fig:5_2024}).  For instance, Neural-UTCI estimates higher values than Polynomial-UTCI in high-elevation areas (Fig. \ref{fig:5_2024}a). This is consistent with Neural-UTCI's improved approximation accuracy under cold and high-wind conditions (Fig. \ref{utci_perf}). The reliability of Neural-UTCI in cold conditions could allow for the development of thermal stress estimates in mountainous environments and UTCI based weather-health warnings for alpine sports.

Additionally, Neural-UTCI estimates systematically lower UTCI in north-eastern North America and in Russia (Fig. \ref{fig:5_2024}a). The variability of differences between approximators is greatest at northern latitudes (Fig. \ref{fig:5_2024}b), driven by strong seasonal contrasts: while the methods show close agreement during summer months, Polynomial-UTCI systematically estimates higher UTCI during the winter (Supplementary Figure S8). Consequently, Neural-UTCI estimates up to 400 additional hours of \textit{very strong cold stress} per grid cell as compared to Polynomial-UTCI. This discrepancy has implications for analyses of human exposure, energy demand, and productivity.

Differences in heat stress exposure are also spatially heterogeneous and vary with regional climate characteristics (Fig. \ref{fig:5_2024}d). In arid desert regions with low humidity, such as the Great Victoria Desert in Australia and the southern Algerian Sahara, Neural-UTCI estimates fewer hours of \textit{very strong heat stress} (Fig. \ref{fig:5_2024}d), despite similar mean annual UTCI (Fig. \ref{fig:5_2024}a).  Humid regions such as Indonesia show a mean UTCI difference of around 1°C  (Fig. \ref{fig:5_2024}a) and a grid cell increase of up to 300 hours of \textit{very strong heat stress} exposure in 2024 using Neural-UTCI (Fig. \ref{fig:5_2024}d). Across regions experiencing \textit{extreme heat stress}, Neural-UTCI estimates a median 13\% increase in annual exposure duration, with local increases of up to 800\% (Fig. \ref{fig:5_2024}f). The contrasting patterns between arid and humid regions are consistent with the approximators' differing representation of humidity effects (Fig. \ref{3}). The largest changes in total 2024 \textit{extreme heat stress} exposure using Neural-UTCI occurred in places that are already prone to dangerous heat levels such as Bolivia, Egypt, Iran, India, and Thailand (Supplementary Figure S9).

Overall, Neural-UTCI estimates greater exposure to both  \textit{extreme heat} and \textit{extreme cold stress} across regions (Fig. \ref{fig:5_2024}e–f), demonstrating that relatively small mean UTCI differences can translate into substantial differences in extreme thermal-stress exposure. 

\section{Conclusions}\label{sec12} 

UTCI is a widely used feels-like temperature index but the underlying physiological model is too computationally expensive for operational use \cite{Freitas2017}.  Polynomial-UTCI is an approximation of the full coupled Fiala-UTCI model that is routinely used in research and for feels-like temperature forecasting.  Polynomial-UTCI is often conflated with Fiala-UTCI with the approximation error considered negligible \cite{Pappenberger2015}-- which we dispute here.  

We present Neural-UTCI as an open-source drop-in replacement, reducing approximation error by 87\% compared to Polynomial-UTCI at a reduced computational cost for batched operational calculations. With Neural-UTCI the approximator is no longer the dominant driver of UTCI error.  We highlight that the physiological heat–humidity relationship established in the underlying Fiala model is faithfully captured by the improved UTCI representation, which is especially important given the emergence of uncompensable humid-heat events \cite{Raymond2020}. We find that differences introduced by UTCI approximation can lead to meaningful downstream classification errors when the index is used for categorical heat stress estimation. Considering the increased incidence of heat-stress days in Europe \cite{Emerton2026}, a fixed thermal-stress misclassification rate corresponds to a growing absolute number of missed dangerous days, potentially impacting warning issued during deadly extreme weather. Neural-UTCI reduces misclassification by 68\% compared to Polynomial-UTCI, with the largest improvements at extreme temperatures. Neural-UTCI is therefore a timely methodological improvement for both heat-health forecasting and climate impact research. 

While Neural-UTCI is more accurate than the current operational procedure, it can only be as accurate as the underlying model itself. The fidelity of Neural-UTCI is further limited by its training data: for instance, the training data released by Bröde et al. (2012) \cite{Brode2012} does not capture typical correlations between the input conditions, which often co-vary. This impacts modeled statistical relationships in both approximators. Additionally, the test data have no cases of extreme heat stress because they are sampled from earth system model control simulations for 1971-80 \cite{Brode2012}. As extreme heat events become more frequent and severe, historical test data are increasingly unrepresentative of the conditions under which UTCI is applied.  Updated reference simulations with the proprietary Fiala-UTCI model spanning present-day and emerging extreme conditions would improve UTCI approximation accuracy and would allow a better understanding of where the models fail. 

Neural-UTCI uses the available Fiala-UTCI simulations to reduce a substantial source of error from operational UTCI calculation while retaining the computational efficiency required for large-scale applications. More accurate feels-like temperature estimation can enable a better understanding of emerging thermal-stress patterns through time and across the globe for targeted mitigation of the deadly effects of  extreme weather.

\section{Funding}
B.P. was supported by the UKRI Natural Environment Research Council, with award no. NE/S007474/1.
M.K. acknowledges funding from the Natural Environment Research Council under grant number UKRI191.
S.W.K. was funded by a Minderoo Foundation grant on Lethal Heat to L.S.

\section{Roles}
\textbf{B.P.} contributed to conceptualization, data curation, formal analysis, investigation, methodology, software, visualization, validation, writing – original draft, and writing – review \& editing.
\newline
\textbf{L.S.} contributed to methodology, supervision, and writing – review \& editing.
\newline
\textbf{M.K.} contributed to methodology, software, supervision, and writing – review \& editing.
\newline
\textbf{T.T.} contributed to methodology and writing – review \& editing.
\newline
\textbf{S.W.K.} contributed to writing – review \& editing.

\section{Supplementary Information}
This article contains a supplementary information document.

\putbib[references]
\end{bibunit}

\clearpage
\begin{bibunit}[sn-nature]

\sisetup{output-exponent-marker=\ensuremath{\mathrm{e}}, detect-mode, group-separator={,}}
\newcolumntype{Y}{>{\raggedright\arraybackslash}X} 
\singlespacing
\renewcommand{\figurename}{Supplementary Figure}
\renewcommand{\tablename}{Supplementary Table}

\setcounter{figure}{0}
\begin{center}
\Large \textbf{Supplementary Information} \\
\vspace{0.5em}

\large \textit{A neural network-based Universal Thermal Climate Index for reliable global thermal-stress classification across extreme weather}\\
\vspace{1.0em}

\large
Bikem Pastine$^{1}$, Milan Klöwer$^{2}$, Tianning Tang$^{3}$, Sarah Wilson Kemsley$^{1}$, Louise Slater$^{1}$ \\
\vspace{0.5em}

\small
$^{1}$School of Geography and the Environment, University of Oxford, Oxford, United Kingdom \\
$^{2}$Atmospheric, Oceanic and Planetary Physics, University of Oxford, Oxford, United Kingdom \\
$^{3}$Department of Mechanical and Aerospace Engineering, University of Manchester, United Kingdom \\

\vspace{0.5em}
Correspondence: bikem.pastine@ouce.ox.ac.uk
\end{center}

\vspace{1.5em}


\section{Supplementary Methods}

\subsection{Universal Thermal Climate Index }\label{UTCI_deff}
UTCI is calculated using four meteorological inputs as an offset $F$ from the 2-meter air temperature $T_a$ \cite{Brode2012}

\begin{equation}
     \text{UTCI}(T_a, T_r, v_a, r_H) = T_a + F(T_a, T_r, v_a, r_H)
\label{eq1}
\end{equation}
\\

where $T_r$ is mean radiant temperature (°C), $v_a$ is 10-meter wind speed magnitude (m/s), and $r_H$ is relative humidity (\%). 

The Fiala multi-node physiological model estimates thermo-physiological responses to reference conditions based on 17 physiological responses. Environmental conditions that produce physiological responses similar to those under the reference conditions are assigned the corresponding air-temperature equivalent using the coupled UTCI, Fiala and clothing models, with equivalence determined by principal component analysis.

The reference conditions are a wind speed of 0.5 m/s, relative humidity of 50\%, and equal mean radiant temperature to air temperature, assuming a walking speed of 4 km/h and a metabolic heat production of 135 \si{W m^{-2}}. 

The UTCI operational procedure further defines thermal stress categories based on the onset of physiological symptoms \cite{Brode2012} (see Table 1).

\subsection{UTCI-Fiala Dataset descriptions}

The 'Grid data' used to train Polynomial-UTCI and the models built in this paper is sampled as follows:

\vspace{0.5em}
\begin{quote}
\begin{itemize}
    \item 2-meter air temperature: $-50^\circ\mathrm{C} \le T_a \le +50^\circ\mathrm{C}$ (1\,\textdegree{}C increments),
    \item Mean radiant temperature: $-80^\circ\mathrm{C} \le T_r \le +120^\circ\mathrm{C}$ (1\,\textdegree{}C increments),
    \item Humidity: $5\% \le \mathrm{rH} \le 100\%$ (5\% increments),
    \item Wind speed (10\,m above ground level): 
    \\$v_a \in \{0.5,\,0.8,\,1.2,\,1.8,\,2.7,\,4.0,\,6.0,\,9.0,\,13.5,\,20.2,\,30.3\}\ \mathrm{m\ s^{-1}}$.
\end{itemize}
\end{quote}
\vspace{0.5em}   

The ECHAM4 evaluation data has the following input domain:

\vspace{0.5em} 
\begin{quote}
\begin{itemize}
    \item 2-meter air temperature: $-50^\circ\mathrm{C} \le T_a \le +42^\circ\mathrm{C}$,
    \item Mean radiant temperature: $-17^\circ\mathrm{C} \le T_r \le +50^\circ\mathrm{C}$,
    \item Wind speed: $1\ \mathrm{m\ s^{-1}} \le v_a \le 30\ \mathrm{m\ s^{-1}}$,
    \item Humidity: $5\% \le \mathrm{rH} \le 100\%$.
\end{itemize}
\end{quote}
\vspace{0.3em}

\subsection{ERA5 Preprocessing}

We use the following data for global and regional UTCI approximator comparisons:
\newline
\newline
The following variables are used:
\begin{quote}
\begin{itemize}
            \item 2-meter surface temperature (ERA5 reanalysis, hourly, single level)
            \item 2-meter dew point temperature (ERA5 reanalysis, hourly, single level)
            \item 10m v- and u-components of wind (ERA5 reanalysis, hourly, single level)
            \item Mean Radiant Temperature (ERA5-HEAT)
            \item  UTCI (ERA5-HEAT) \\
\end{itemize}
\end{quote}

These input parameters are processed as follows to run the neural network approximator on the ERA5 data.

\begin{enumerate}
    \item 2 meter surface temperatures, mean radiant temperature, and UTCI are converted from Kelvin to Celsius.

    \item Dew point temperature is converted to relative humidity (\%) using the Clausius–Clapeyron relation \cite{Alduchov1996Magnus}.
    
\[
\mathrm{RH}
= 100
\exp\left[
\frac{L_v}{R_v}
\left(
\frac{1}{T_a}
-
\frac{1}{T_d}
\right)
\right],
\]
where RH is relative humidity (\%),  $T_a$  and $T_d$ are air temperature and dewpoint temperature in Kelvin, $L_v$ is the latent heat of vaporization ($2.26\times10^6$ J/kg), and $R_v$ is the specific gas constant for water vapor (461.5 J/kg·K)
 
\item 10m v- and u-component of wind converted to 10m wind speed by:
\[
V = \sqrt{u^2 + v^2},
\]
where V is 10m wind speed in m/s, u is the u- component, and v is the v- component in m/s. 
\end{enumerate}

\subsection{Verification metrics}
Root Mean Square Error (RMSE) is used to assess the performance of the models and is given by:
\[
\mathrm{RMSE} = \sqrt{\frac{1}{n} \sum_{i=1}^{n} (y_i - \hat{y}_i)^{2}} 
\label{eq2}
\]
Where n is the number of observations, $y_i$ is the Fiala-UTCI model value, and $\hat{y}_i$ is the predicted value by the approximating function. Lower RMSE values indicate better model performance. RMSE is sensitive to outliers and thus gives insight into how large the errors of a model can be. It is the evaluation metric used in Bröde et al. (2012) \cite{Brode2012}.

The Mean Error (ME) is used to evaluate the bias in the models and is given by:
\[
\mathrm{ME} = \frac{1}{n} \sum_{i=1}^{n} (y_i - \hat{y}_i)
\label{eq3}
\]
A ME closer to zero indicates an unbiased model. Positive values indicate systematic underprediction, and negative values indicate overprediction.

The Mean Absolute Error (MAE) measures the magnitude of the average error and is defined as:
\[
\mathrm{MAE} = \frac{1}{n} \sum_{i=1}^{n} \lvert y_i - \hat{y}_i \rvert\label{eq4}
\]
The lower the MAE, the better the prediction. This metric helps to compare models, regardless of whether under- or overpredicted.

Maximum Error (MaxE) is the largest error that the model has on the testing dataset and is given by:
\[
\mathrm{MaxE} = \max \lvert y_i - \hat{y}_i \rvert\label{eq5}
\]

The maximum error is unique to the testing dataset and does not indicate global model performance, but it is useful in comparing the worst-case performance between models evaluated on the same data.

\subsection{Extended regression model training methods}
\subsubsection{Representational model comparison}
Models for each candidate method were built and hyperparameters were tuned with Bayesian hyperparameter optimization using the python \textit{optuna} package \cite{optuna_2019}. Input data was normalized by fitting the transformation to the training data and applying it to the validation and testing datasets. Models were optimized to minimize RMSE, and validation and training performance were monitored for overfitting. Where overfitting occurred, regularization strength and other hyperparameter tuning ranges were adjusted to improve generalization. No early stopping was employed, except for the neural network, all trials were executed to completion. The modeling methods investigated were support vector machine, random forest regression, XGBoost, and artificial neural networks.  The search space and selected hyperparameters can be found in Supplementary Table S\ref{tab:hyperparams}.

Polynomial ordinary least squares regression models with interaction terms of orders 1-8 were compared, along with ridge regularized ($\alpha$ =10) versions of the models. The unregularized sixth-order polynomial was identical to the operational UTCI emulator developed by Bröde et al. (2012) \cite{Brode2012}, except that 10\% of the training grid data was withheld for validation. The results of this analysis can be found in SM2. 

\subsubsection{Neural Network}
Feed-forward Neural Networks are well suited to capture regression relationships that are complex and nonlinear in high dimensional input space; that ordinary least squares linear regression can often struggle with. UTCI is one such system, where the relative weight of each of the input parameters differs depending on the value of the other input parameters, and dynamics can be complex. As such, NN is a well-suited method to address the need for an improved UTCI emulator. 

A feed-forward NN model was compared to other ML models as described above and was found to be the best performing method out of the models tested. The NN architecture was refined to maximize the performance of the final operational procedure. This was done in a systematic manner, whereby an optimal architecture was found, and then other hyperparameters were tuned to minimize loss calculated on the validation dataset.

The investigatory NN models were trained on scaled training variables to increase stability. \textit{Ta }and \textit{Tr-Ta} were standardized using \textit{StandardScaler} following convention (mean = 0 and std =1). The input parameters \textit{pa }and \textit{va }were min-max scaled due to their heavy-tailed distributions and as min-max standardization for \textit{pa }and \textit{va }was found to improve model performance. These scalers were fit to the training data and were then applied to the validation and the testing dataset. The scaling was exported for prediction.

To identify an optimal NN architecture, ten representative architectures were systematically tested. Models were trained using \textit{scikit-learn}’s \textit{MLPRegressor}. Hyperparameters were set to default values (\textit{\(\alpha\) = 0.001, loss function = MSE, Activation function = ReLU}). Each model was trained for 100 epochs using the Adam optimizer and warm-start functionality. Loss curves with MSE were investigated to identify overfitting, and no sign of overfitting in any of the architectures was observed. 

As a first step in identifying an appropriate NN architecture, the best performing NN model size was determined. The minimum total trainable parameter number was set to 210 to match the coefficients of Polynomial-UTCI, and the maximum was capped at 10\% of the training data size to prevent overfitting. Three sizes were tested as detailed in the supplementary materials. The ‘large’ NN model was selected as the best performing model based on its lower validation MAE, MSE, and RMSE, as compared to the other architectures.

Next, the number of hidden layers and structure of nodes was investigated following the same procedure. As such, the architecture and hyperparameters were optimized sequentially rather than through a full joint search to reduce the computational burden. The best performing model was determined to be three-hidden-layer NN with a decreasing number of nodes (4, 79, 75, 39, 1) based on all three criteria. See Table 2 in the supplementary materials for results of this model comparison.

Finally, to design the optimal model configuration, hyperparameter tuning was conducted on the selected model architecture. Exhaustive grid search was used to investigate 96 combinations of learning rate, regularization, batch size, and loss function. Values investigated for each of these hyperparameters can be found in Supplementary Table S\ref{tab:hyperparams}. Models were built using the \textit{pytorch} package. Model performance was compared based on validation RMSE loss. 

The Adam optimizer and ReLU activation function were selected and omitted from tuning. The ReLU activation function is appropriate for the UTCI emulation task as it allows for the model flexibility appropriate for the nonlinear behavior of UTCI \cite{Dubey2022}. The Adam optimizer was selected to leverage its fast convergence and stable results.

The close-optimal set of hyperparameters was found to be a learning rate of 0.001, a MSE loss function, no regularization, and a batch size of 16 observations. The final model was trained for 139 epochs as this is when there is no improvement in model performance with a patience of 25 epochs and a minimum delta of 0.0001.

\subsection{Error Decomposition}

We decomposed the total error in UTCI into two components:
\begin{enumerate}
    \item Error associated with the Fiala-UTCI multi-node thermophysiological model.
    \item Error introduced by the approximation of Fiala-UTCI.
\end{enumerate}

Let $UTCI_{\mathrm{true}}$ be the unobserved true equivalent temperature:
\begin{align}
\varepsilon_{\mathrm{Fiala}} 
&= UTCI_{\mathrm{Fiala}} - UTCI_{\mathrm{true}}, \\
\varepsilon_{\mathrm{Approx}} 
&= UTCI_{\mathrm{Approx}} - UTCI_{\mathrm{Fiala}}.
\end{align}
Where $\varepsilon$ represents the error. The total error of the approximated UTCI relative to the true value is therefore:
\begin{equation}
\varepsilon_{\mathrm{Total}} 
= \varepsilon_{\mathrm{Fiala}} + \varepsilon_{\mathrm{Approx}}.
\end{equation}

Using the variance sum law:
\begin{equation}
\mathrm{Var}(\varepsilon_{\mathrm{Total}})
=
\mathrm{Var}(\varepsilon_{\mathrm{Fiala}})
+
\mathrm{Var}(\varepsilon_{\mathrm{Approx}})
+
2\,\mathrm{Cov}(\varepsilon_{\mathrm{Fiala}}, \varepsilon_{\mathrm{Approx}}).
\end{equation}

We assume unbiased, normally distributed errors and let the mean squared error equal the variance. To the knowledge of these authors, the RMSE of  Fiala-UTCI has not been reported in the literature. The error associated with outputs of the Fiala-UTCI multi-node model coupled with the clothing model are reported: with dynamic thermal sensation having a RMSE of 1.04°C averaged over simulated exposure times \cite{Broede2012Brazil}, skin temperature a RMSE of 1.35°C \cite{Psikuta2012}, and core temperature a RMSE of 0.32°C \cite{Psikuta2012}. Dynamic thermal sensation is an emergent property of the other components of UTCI and so the error reported for this variable is assumed to be similar to the error of Fiala-UTCI as a whole. The additional approximation error was bounded between zero and the reported RMSE of Polynomial-UTCI across the full wind range.

We compute total error in UTCI in Supplementary Figure S2 under:
\begin{equation}
\rho = -1,\, 0,\, +1,
\end{equation}
where $\rho$ is the correlation coefficient between the two error components.

We assume independence between $\varepsilon_{\mathrm{Fiala}}$ and $\varepsilon_{\mathrm{Approx}}$ to calculate the additional increase in mean squared error from the approximator.

\newpage
\section{Supplementary Figures}

\begin{figure}[H]
    \centering
    \includegraphics[width=1\textwidth]{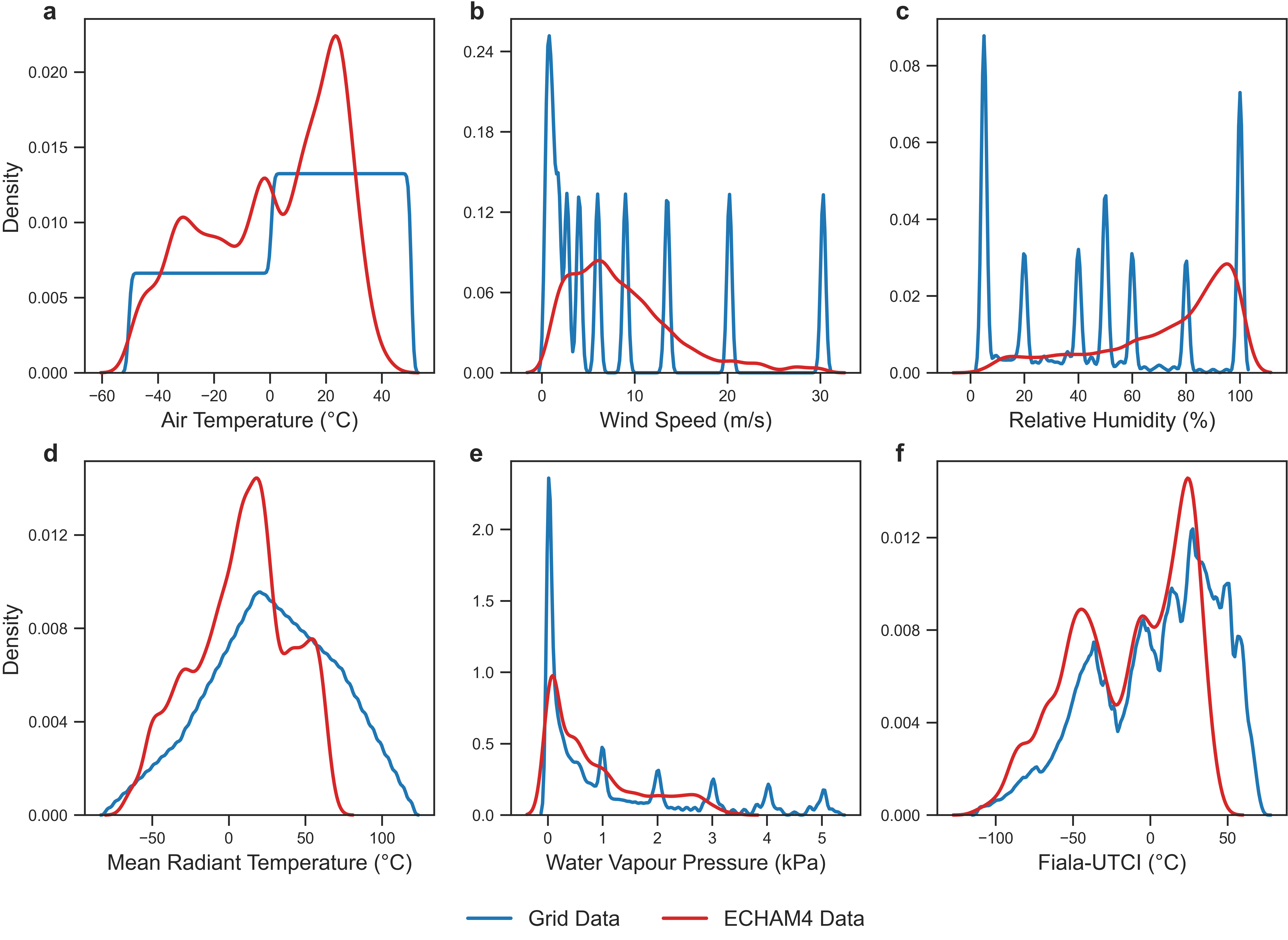}
    \caption{\textbf{Grid data and ECHAM4 dataset distributions.} Grid data (blue) is used for model training and validation and the ECHAM4 data (red) is used for testing. Subplots show \textbf{(a)} air temperature (\textdegree{}C), \textbf{(b)} wind speed (m/s), \textbf{(c)} relative humidity (\%), \textbf{(d)} mean radiant temperature (\textdegree{}C), \textbf{(e)} water vapor pressure (kPa), and the resulting Fiala-UTCI value (\textdegree{}C). The y-axis shows the density of the value given on the x axis.
}    \label{fig:supp1}
\end{figure}
\newpage

\begin{figure}[H]
    \centering
    \includegraphics[width=1\textwidth]{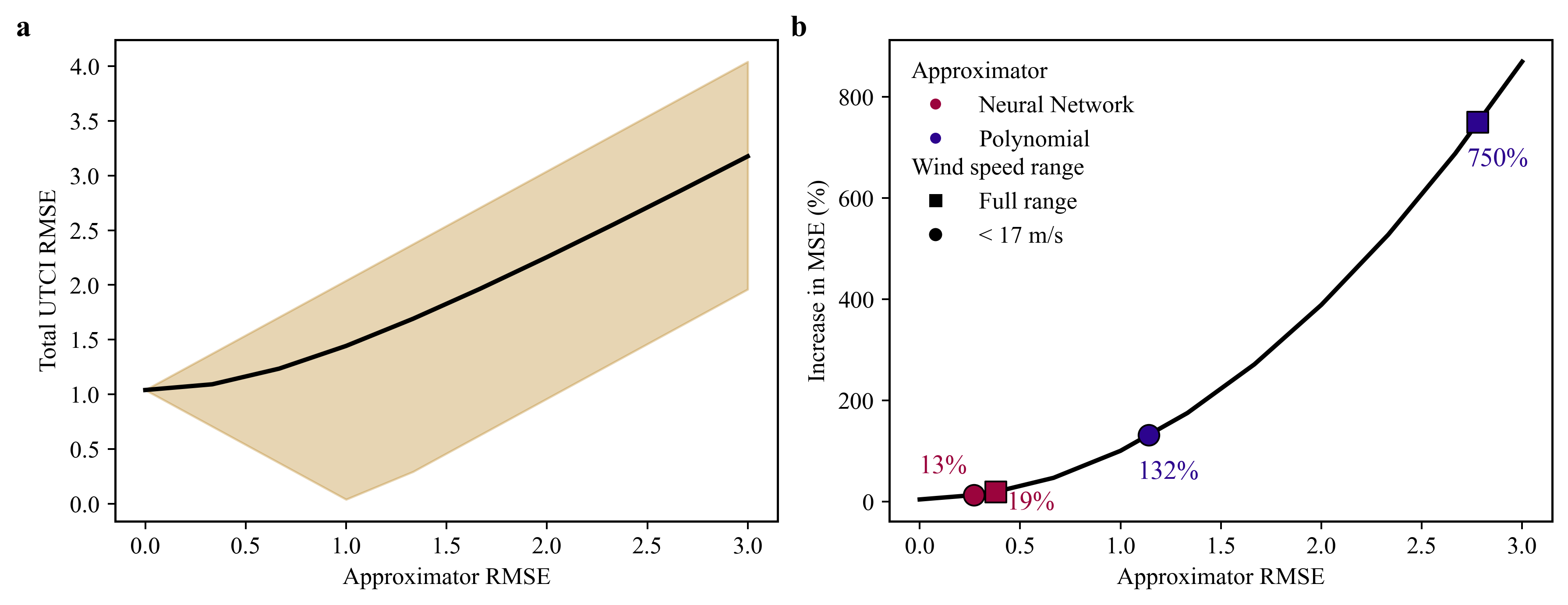}
    \caption{\textbf{Error decomposition of total UTCI error from the Fiala model and the approximators.} Subplots show \textbf{(a)} the theoretical total UTCI error, defined as the combined error from the Fiala-UTCI model and the approximation, across a range of approximation RMSE values. The black line represents the estimated total error assuming no correlation between the two error sources, while the yellow shaded region indicates the range of possible total errors under perfect positive or negative correlation. \textbf{(b)} The increase in total mean squared error (MSE) relative to the Fiala-UTCI model for different approximator RMSE values. Points show the percentage increase in MSE for Neural-UTCI (red) and Polynomial-UTCI (blue), evaluated with wind speeds restricted to $<$17 m/s (circles) and across the full UTCI range (squares).}
  \label{fig:supp2}
\end{figure}
\newpage

\begin{figure}[H]
    \centering
    \includegraphics[width=1\textwidth]{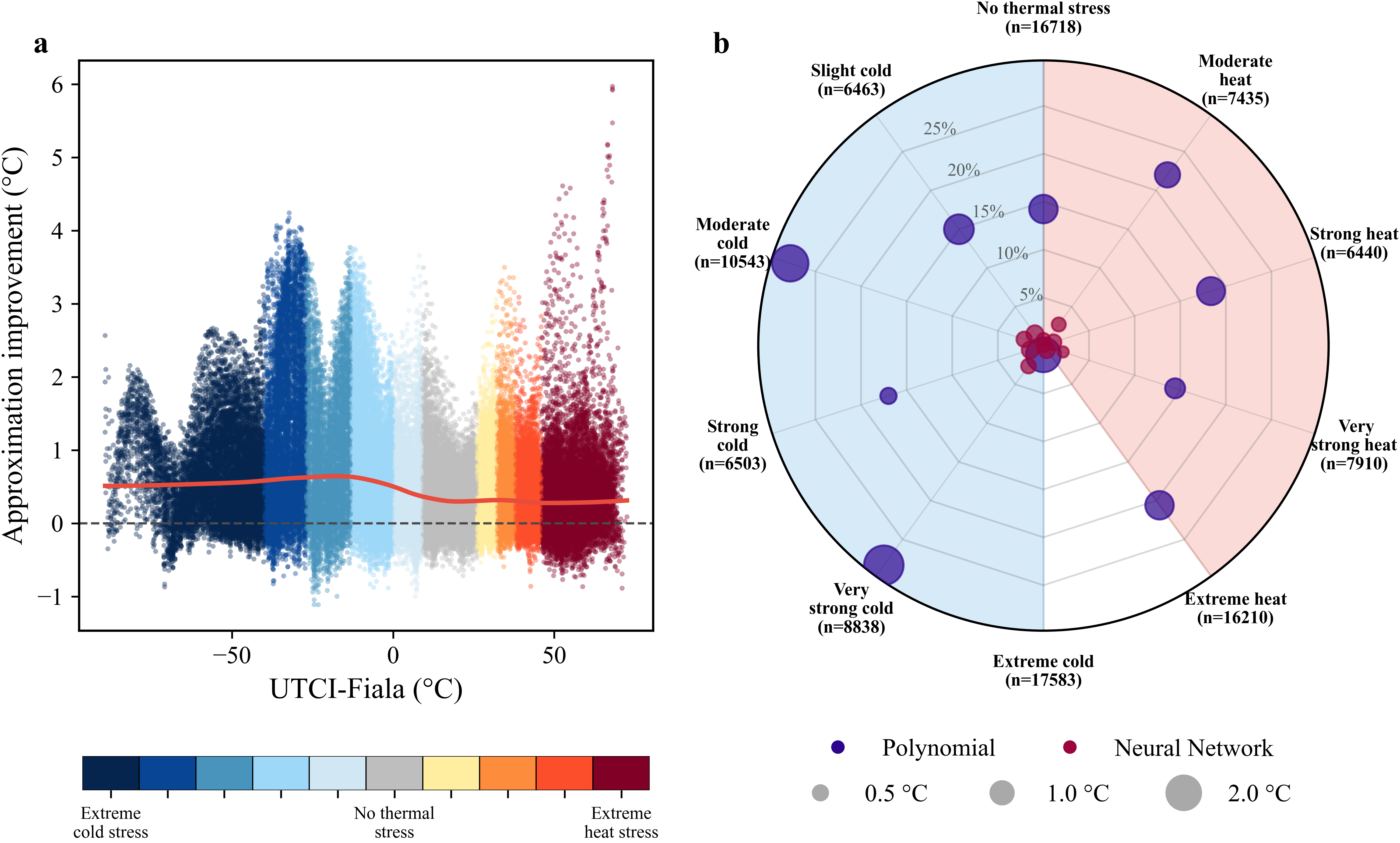}
    \caption{\textbf{Thermal stress miscategorisation analysis on the training dataset.} Performance of Polynomial-UTCI and Neural-UTCI for approximating the
Fiala-UTCI thermal stress categories. \textbf{(a) }Approximation improvement of Neural-UTCI over Polynomial-UTCI for Fiala-UTCI values where wind speed is under 17 m/s. Approximation improvement is defined as the absolute value of Polynomial-UTCI minus the absolute value of Neural-UTCI approximation error for each data point. The red line is a LOESS smoothed regression line, with the span of 40\%. Points are colored by Fiala-UTCI thermal stress categories.\textbf{ (b)} Thermal stress categorization error using Neural-UTCI (red circles) and Polynomial-UTCI (blue circles) as compared to the Fiala-UTCI category. Each sector of the circle represents a thermal stress category, where light red shaded sectors represent heat stress categories and light blue sectors represent cold stress. The distance of a point from the center of the plot represents the percentage of the training data in that category that is miscategorised by the approximation model according to the Fiala-UTCI. The size of the point represents the average absolute approximation error (°C) of the miscategorized points. The labels of the sectors include the number of data points (n) in each thermal stress category.}    \label{fig:supp3}
\end{figure}
\newpage

\begin{figure}[H]
    \centering
    \includegraphics[width=0.7\textwidth]{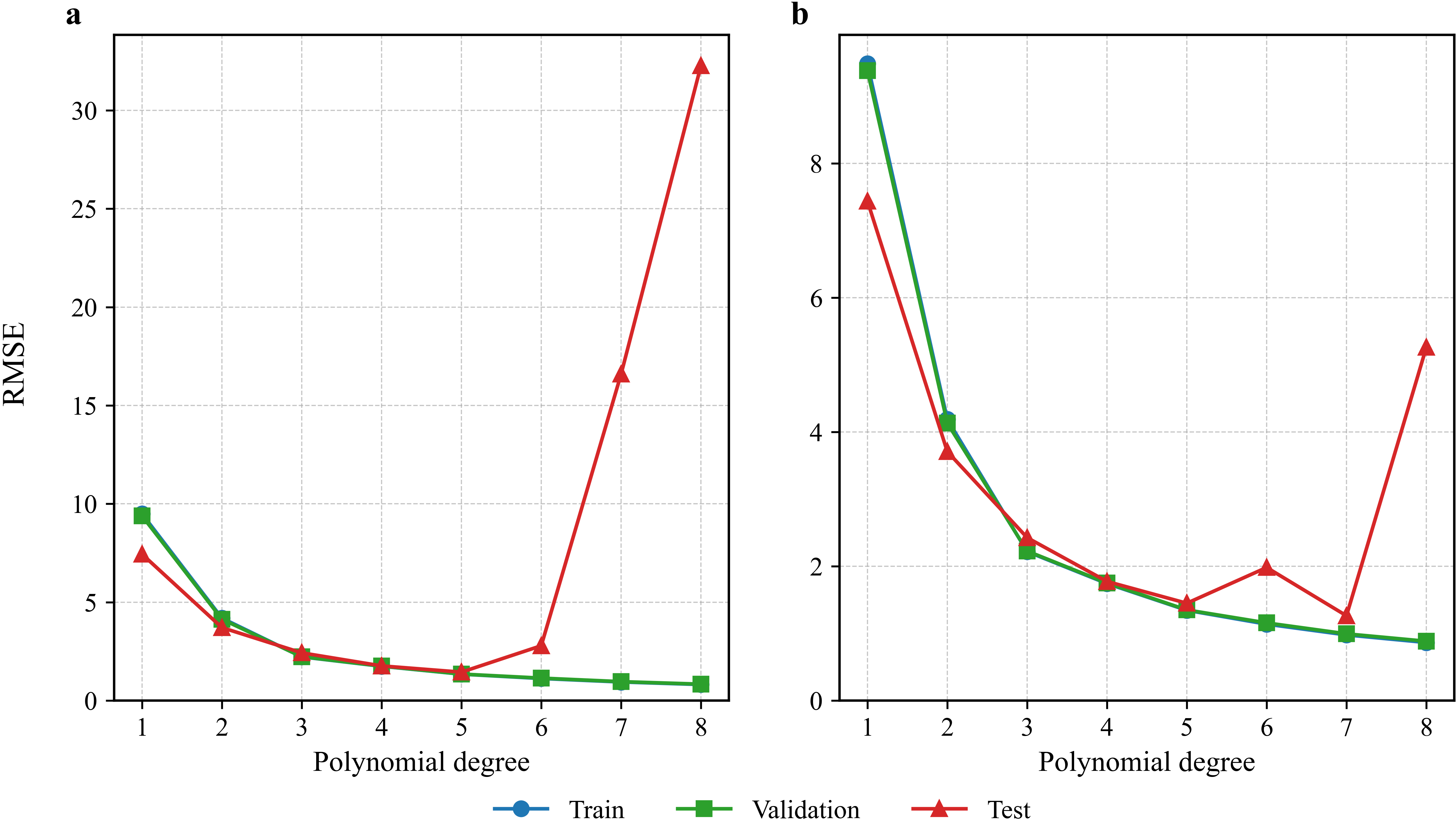}
    \caption{\textbf{Regression model comparison with increasing orders of polynomial magnitude.} Blue line (behind the green line) shows the UTCI RMSE (°C) of the training data, green line shows the model performance on the validation data, and the red line shows performance on the testing data. \textbf{(a)} Ordinary least squares regression, as described in the 2012 operational procedure. The polynomial with a 6\textsuperscript{th} order is equivalent to Polynomial-UTCI as presented in the Bröde et al. (2012) paper, other than being trained on  90\%  of the Grid dataset. \textbf{(b)} Ridge regularized ordinary least squares regression (with alpha set to 10).}
    \label{fig:supp4}
\end{figure}
\newpage
\begin{figure}[H]
    \centering
    \includegraphics[width=0.7\textwidth]{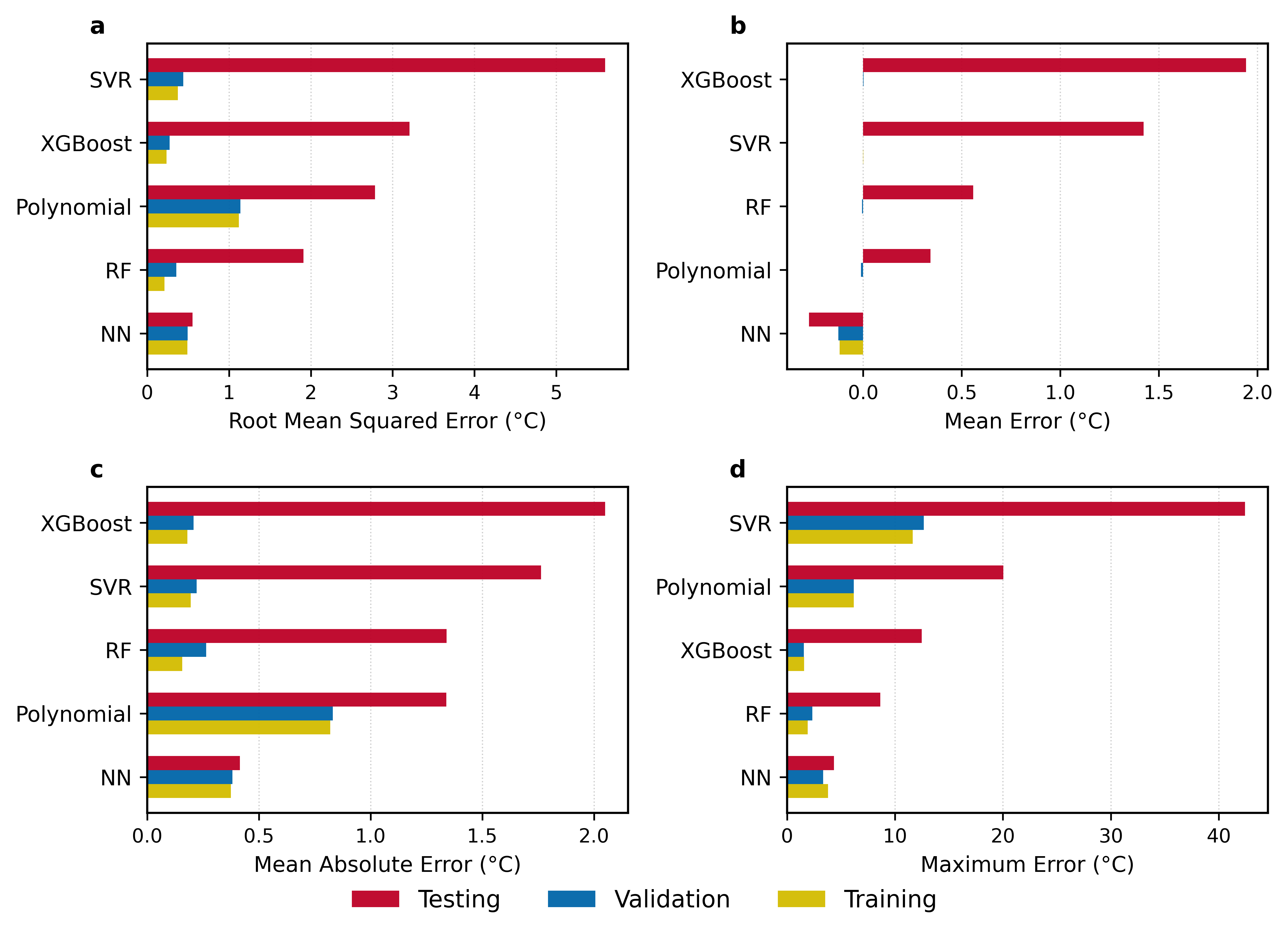}
    \caption{\textbf{Regression model performance comparison of five regression methods evaluated using four error metrics.} Model performance evaluated using four error metrics: Root Mean Squared Error (\textbf{a}), Mean Error (\textbf{b}), Mean Absolute Error (\textbf{c}), and Maximum Error (\textbf{d}). Each subplot shows the model performance on the independent ECHAM4 testing set (red), the validation set (blue), and the training set (yellow bars). Models are ordered by performance measured by the testing error. Regression models include the sixth-order operational polynomial (Polynomial-UTCI), Support Vector Regression (SVR), Random Forest (RF), XGBoost, and an Artificial Neural Network (NN). }
    \label{fig:supp5}
\end{figure}
\newpage
\begin{figure}[H]
    \centering
    \includegraphics[width=0.7\textwidth]{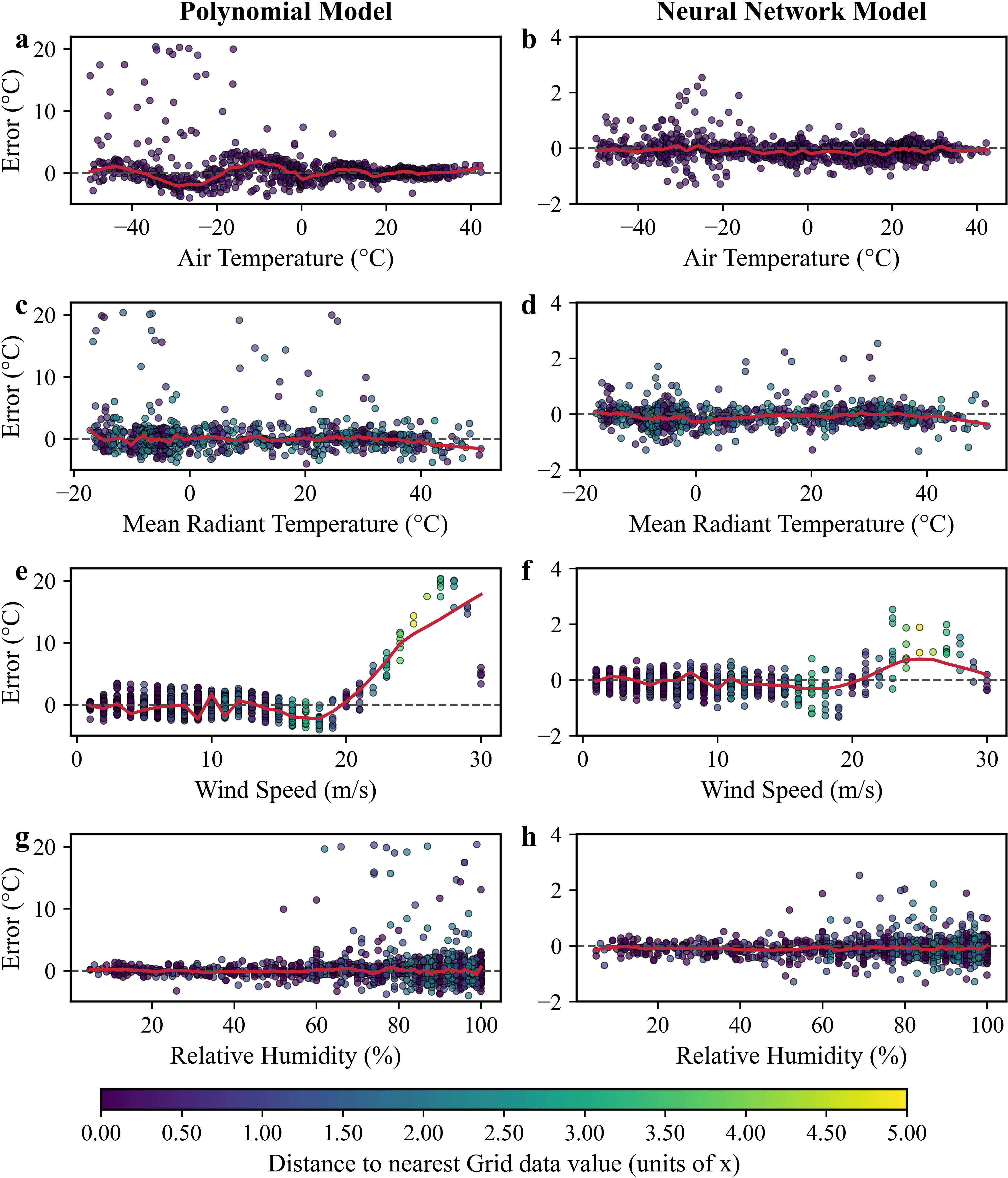}
    \caption{\textbf{Polynomial-UTCI and Neural-UTCI error structure by input parameter. } The error (\textdegree{}C ) for Polynomial-UTCI and Neural-UTCI is shown for the ECHAM4 testing dataset for each input parameter. Each row corresponds to one input parameter with \textbf{(a-b)} showing air temperature (\textdegree{}C), \textbf{(c-d)} mean radiant temperature (\textdegree{}C), \textbf{(e-f)} wind speed (m/s), and \textbf{(g-h)} showing relative humidity (\%).  Error is calculated as UTCI calculated by the approximating model minus the Fiala-UTCI UTCI. Scatter points are colored by the distance of the ECHAM4 input data value to the nearest Grid data sampling point, in unnormalised units of the input variable. Note the different scales of the y-axis.  The red line shows the LOESS regression line. The gray dashed line shows zero error.
}
    \label{fig:supp6}
\end{figure}

\newpage
\begin{figure}[H]
    \centering
    \includegraphics[scale=0.6]{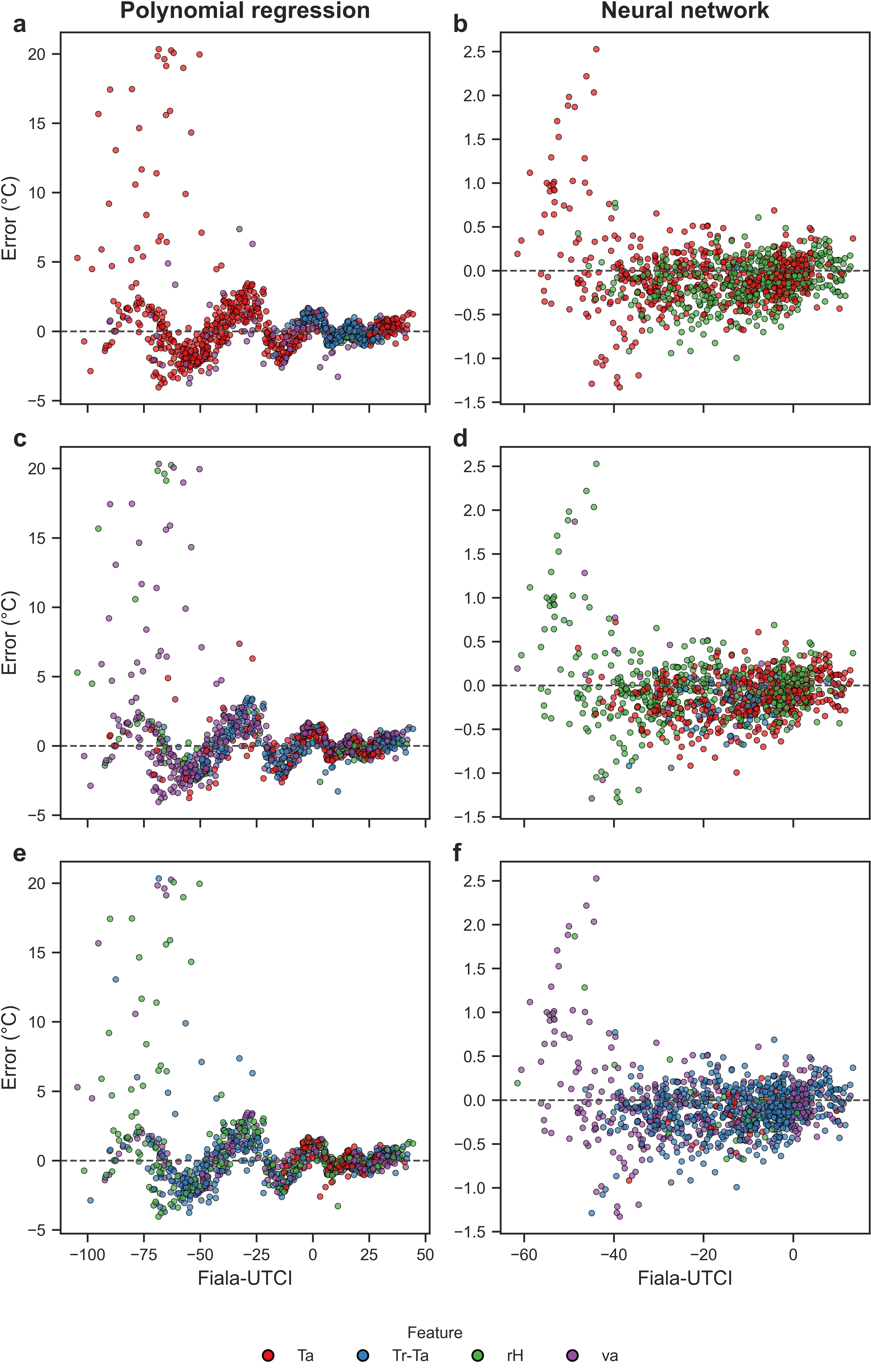}
    \caption{\textbf{SHAP calculated input variable importance for each row of the ECHAM4 testing dataset.} The error (\textdegree{}C), defined as the approximating model UTCI minus Fiala-UTCI. Note the difference in y-axis ranges. Each row shows the \textbf{(a-b)} 1st, \textbf{(c-d)} 2nd, and \textbf{(e-f)} 3rd most influential feature for UTCI calculation in that instance. Data is colored by input parameter that is found to be most influential with red being air temperature (\textdegree{}C) (\textit{Ta}), blue being mean radiant temperature (\textdegree{}C) (\textit{Tr-Ta}), green being relative humidity (\%) (\textit{rH}), and purple being wind speed (m/s) (\textit{va}). The dashed gray line shows zero error. 
}
    \label{fig:supp7}
    \end{figure}

\newpage

\begin{figure}[H]
    \centering
    \includegraphics[scale=0.5]{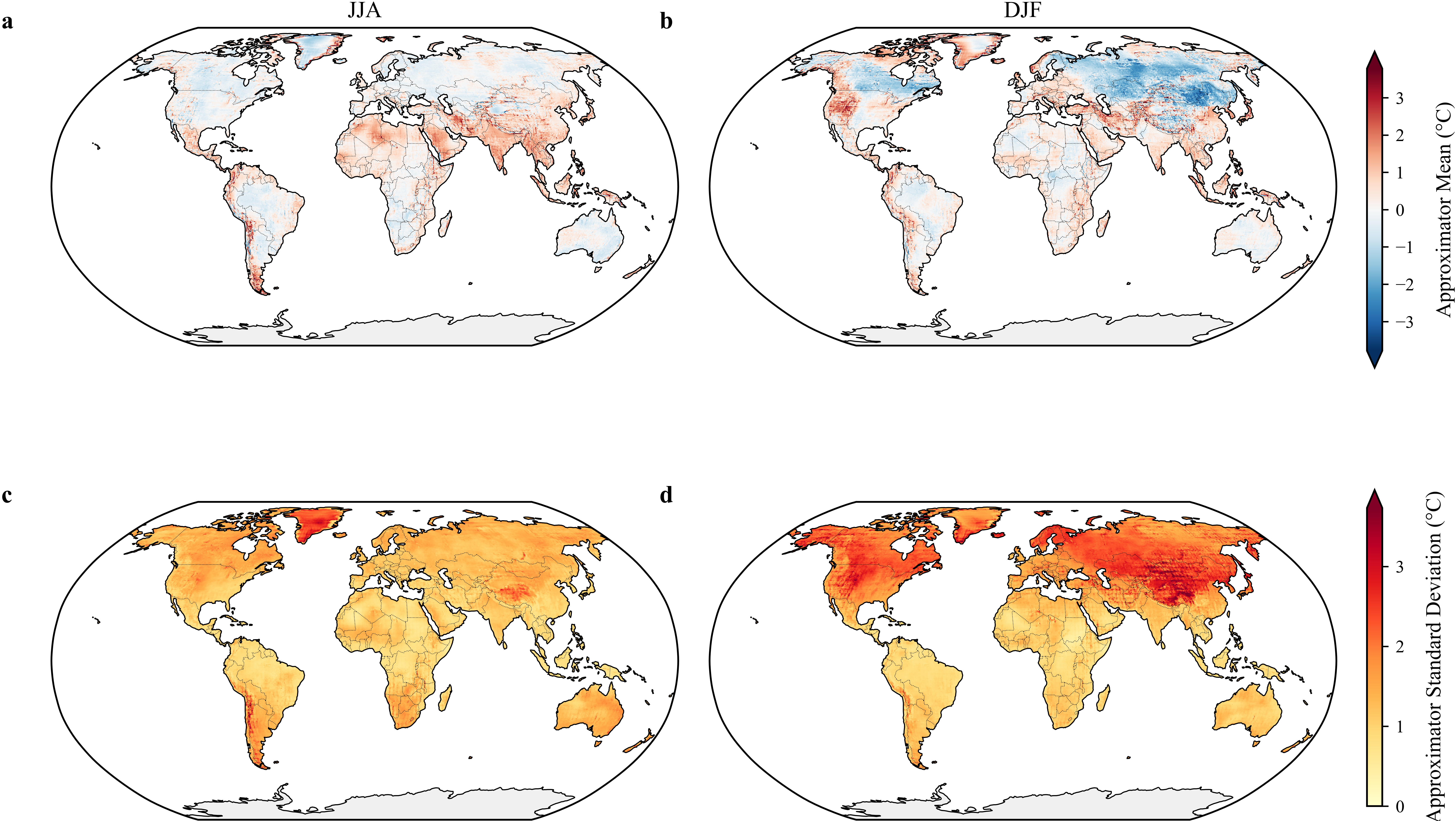}
    \caption{\textbf{Seasonal patterns of approximator difference in UTCI calculated using ERA5 2024 data.} The average  differences in UTCI across seasons is analyzed with mean differences across June-August and December and January-February \textbf{(a-b)}. Variation in UTCI difference is shown between methods with standard deviation \textbf{(c-d).}
}
    \label{fig:supp8}
    \end{figure}

\newpage

\begin{figure}[H]
    \centering
    \includegraphics[scale=0.5]{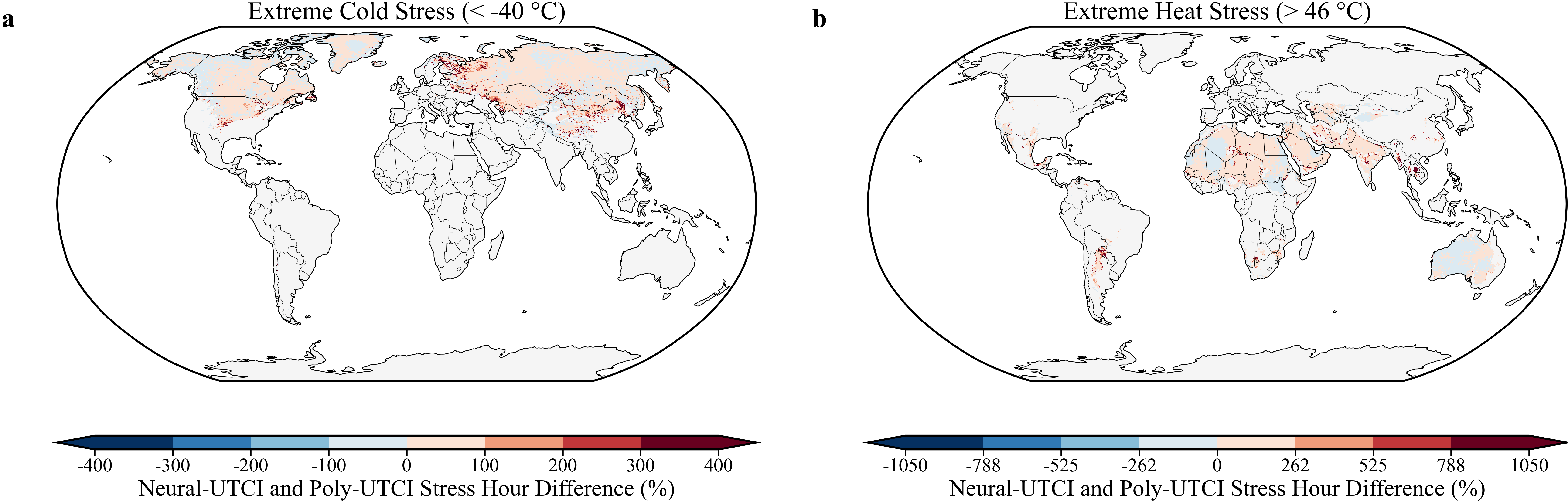}
    \caption{\textbf{Extreme Stress Hour differences in Polynomial-UTCI and Neural-UTCI for ERA5 2024 data.} For \textbf{(a)} extreme cold stress and \textbf{(b)} extreme heat stress. 
}
    \label{fig:supp9}
    \end{figure}

\newpage
\section{Supplementary Tables}

\begin{table}[htbp]
\centering
\caption{UTCI thermal ranges for thermal stress categories. \cite{Brode2012} }
\label{utci_categories}
\begin{tabular}{ll}
\hline
\textbf{UTCI Range (\textdegree{}C)} & \textbf{Thermal Stress Category} \\
\hline
$>$ 46.0 & Extreme heat stress \\
38.0 -- 46.0 & Very strong heat stress \\
32.0 -- 38.0 & Strong heat stress \\
26.0 -- 32.0 & Moderate heat stress \\
9.0 -- 26.0 & No thermal stress \\
0.0 -- 9.0 & Slight cold stress \\
--13.0 -- 0.0 & Moderate cold stress \\
--27.0 -- --13.0 & Strong cold stress \\
--40.0 -- --27.0 & Very strong cold stress \\
$<$ --40.0 & Extreme cold stress \\
\hline
\end{tabular}
\end{table}

\begin{table}[h]
\centering
\caption{\textbf{Valid input ranges for UTCI calculations for Neural-UTCI and Polynomial-UTCI.} Ranges restricted by the training data range released by Br\"{o}de et al. (2012).}
\begin{tabular}{lcc}
\toprule
\textbf{Variable} & \textbf{Neural-UTCI}& \textbf{Polynomial-UTCI}\\
\midrule
2-m Air Temperature ($T_a$) & $-50$\textdegree{}C to $+50$\textdegree{}C & $-50$\textdegree{}C to $+50$\textdegree{}C \\
Mean Radiant Temperature ($T_r$) & $-80$\textdegree{}C to $+120$\textdegree{}C & $-80$\textdegree{}C to $+120$\textdegree{}C \\
Relative Humidity ($rH$) & 5\% to 100\% & 5\% to 100\% \\
Wind Speed ($v_a$ at 10 m) & 0.5 to 30.3 m\,s$^{-1}$ & 0.5 to 17 m\,s$^{-1}$ \\
\bottomrule
\end{tabular}
\label{tab:utci_ranges}
\end{table}

\pagebreak
\begin{table}[h]
\centering
\caption{\textbf{Categorical approximation performance comparison between Neural-UTCI and Polynomial-UTCI.} RMSE computed on Fiala-UTCI in the testing dataset for each thermal stress category.}
\begin{tabular}{lccc}
\toprule
\textbf{Category} & \textbf{RMSE Neural-UTCI}& \textbf{RMSE Polynomial-UTCI}& \textbf{\% Decrease}\\
\midrule
Extreme cold stress        & 0.52 & 4.85 & 89.18 \\
Very strong cold stress    & 0.32 & 2.03 & 84.31 \\
Strong cold stress         & 0.34 & 1.31 & 74.23 \\
Moderate cold stress       & 0.31 & 1.13 & 72.78 \\
Slight cold stress         & 0.34 & 0.84 & 59.00 \\
No thermal stress          & 0.23 & 0.54 & 57.86 \\
Moderate heat stress       & 0.20 & 0.44 & 54.76 \\
Strong heat stress         & 0.16 & 0.55 & 71.02 \\
Very strong heat stress    & 0.17 & 0.70 & 75.69 \\
\bottomrule
\end{tabular}
\label{tab:rmse_comparison}
\end{table}

\pagebreak

\begin{table}[h]
\centering
\caption{\textbf{Computational speed of UTCI approximation models.} The models are timed over 1 million \textit{float32} input samples, either batched or sequentially looped. The GPU used is a single NVIDIA L40, with 48 GB of memory. Reported uncertainty is given by the standard deviation over 30 $\times$ 1 million sample reruns.}
\begin{tabular}{lcccc}
\toprule
\textbf{Method} & \multicolumn{2}{c}{\textbf{CPU (seconds)}} & \multicolumn{2}{c}{\textbf{GPU (seconds)}} \\
\cmidrule(lr){2-3} \cmidrule(lr){4-5}
 & \textbf{Sequential} & \textbf{Batched} & \textbf{Sequential} & \textbf{Batched} \\
\midrule
Polynomial-UTCI   & 20.64 $\pm$ 0.20 & 0.28 $\pm$ 0.01 & -- & -- \\
Neural-UTCI & 25.38 $\pm$ 0.56 & 0.21 $\pm$ 0.02 & 72.13 $\pm$ 0.40 & \(8.70 \times 10^{-3}\) $\pm$ \(0.06 \times 10^{-3}\) \\
\bottomrule
\end{tabular}
\label{tab:comp_speed}
\end{table}
\pagebreak
\begin{table}[htbp]
\centering
\small
\caption{\textbf{UTCI approximation model hyperparameter search spaces and selected values.} Bayesian random search was performed with \texttt{optuna}.}
\label{tab:hyperparams}
\setlength{\tabcolsep}{6pt}
\begin{tabularx}{\textwidth}{l Y Y Y}
\toprule
\textbf{Model} & \textbf{Hyperparameter} & \textbf{Search space} & \textbf{Selected value} \\
\midrule

\multirow{6}{*}{XGBoost}
 & maximum depth & $[3, 8]$ & 8 \\
 & learning rate ($\eta$) & $[10^{-3},\,3\cdot10^{-1}]$ & 0.153 \\
 & subsample & $[0.5, 1.0]$ & 1.000 \\
 & column sample by tree & $[0.5, 1.0]$ & 0.910 \\
 & L1 regularization ($\alpha$) & $[0, 10]$ & 0.210 \\
 & L2 regularization ($\lambda$) & $[0, 10]$ & 7.416 \\
\midrule

\multirow{5}{*}{SVR}
 & $C$ & $[10^{-2},\,10^{3}]$ & 1.965 \\
 & gamma ($\gamma$) & \{scale, auto, float\} & float — 7.575 \\
 & epsilon ($\epsilon$) & $[10^{-6},\,10^{-2}]$ & $9.15\times10^{-6}$ \\
 & maximum iterations & \{1000, 5000, -1\} & -1 \\
 & kernel & rbf (fixed) & rbf \\
\midrule

\multirow{6}{*}{Random Forest}
 & number of estimators & \{100, 200, 500\} & 200 \\
 & maximum depth & $[3, 20]$ & 20 \\
 & minimum samples per leaf & $[1, 10]$ & 4 \\
 & minimum samples per split & $[2, 20]$ & 7 \\
 & max features & \{sqrt, log2, None\} & log2 \\
 & bootstrap & \{True, False\} & False \\
\midrule

\multirow{8}{*}{Neural Network (NN)}
 & number of layers & $[1, 3]$ & 2 \\
 & units per layer & $[16, 256]$ & 53 \\
 & dropout rate & $[0.0, 0.5]$ & 0.0065 \\
 & learning rate ($\eta$) & $[10^{-4},\,10^{-2}]$ & 0.00139 \\
 & weight decay & $[10^{-6},\,10^{-2}]$ & $7.47\times10^{-4}$ \\
 & batch size & \{64, 128, 256\} & 128 \\
 & activation & \{relu, tanh\} & relu \\
 & early stopping patience & $[5, 10]$ & 7 \\
\bottomrule
\end{tabularx}
\end{table}

\pagebreak

\begin{table}[htbp]
\centering
\caption{\textbf{Architectures and parameter count of the tested NN architectures for manual tuning}. The number of layers and topology tested were selected to have a similar number of parameters estimated as the 'large' architecture. }
\label{tab:architecture}
\begin{tabular}{l l l c}
\hline
\textbf{Category} & \textbf{Model} & \textbf{Shape} & \textbf{Params} \\
\hline
Size   & Small  & (4, 14, 9, 1)       & 215 \\
       & Medium & (4, 30, 28, 1)      & 1,047 \\
       & Large  & (4, 110, 79, 1)     & 9,399 \\
Layers & 1      & (4, 1566, 1)        & 9,397 \\
       & 2      & (4, 110, 79, 1)     & 9,399 \\
       & 3      & (4, 79, 75, 39, 1)  & 9,399 \\
       & 4      & (4, 61, 59, 53, 41, 1) & 9,399 \\
Topology & Increasing. & (4, 7, 65, 132, 1)  & 9,500 \\
         & Decreasing & (4, 79, 75, 39, 1)  & 9,400 \\
         & Even & (4, 67, 67, 67, 1)  & 9,399 \\
\hline
\end{tabular}
\end{table}
\pagebreak

\begin{table}[htbp]
\centering
\caption{\textbf{Model performance metrics for the manually tested NN architectures.} Best performing architectures in each stage are written in bold.}
\label{tab:performance}
\begin{tabular}{l l c c c}
\hline
\textbf{Category} & \textbf{Model} & \textbf{MSE} & \textbf{MAE} & \textbf{RMSE} \\
\hline
Size   & Small        & 0.909 & 0.748 & 0.953 \\
       & Medium       & 0.233 & 0.369 & 0.483 \\
       & \textbf{Large}        & \textbf{0.114} & \textbf{0.248} & \textbf{0.338} \\
Layers & 1 layer      & 0.213 & 0.347 & 0.461 \\
       & 2 layers     & 0.114 & 0.248 & 0.338 \\
       & \textbf{3 layers}     & \textbf{0.091} & \textbf{0.235} & \textbf{0.302} \\
       & 4 layers     & 0.097 & 0.239 & 0.311 \\
Topology & Increasing & 0.213 & 0.359 & 0.461 \\
         & \textbf{Decreasing} & \textbf{0.091} & \textbf{0.235} & \textbf{0.302} \\
         & Even       & 0.097 & 0.244 & 0.312 \\
\hline
\end{tabular}
\end{table}

\pagebreak
\begin{table}[h!]
\centering
\caption{\textbf{ Neural Network hyperparameters explored by grid search.} Hyperparameters are tuned using grid search. A total of 96 combinations were evaluated and the best performing model based on RMSE was selected.  A learning rate of 0.001, a MSE loss function, no regularization, and a batch size of 16 was found to be the optimal combination.}
\begin{tabular}{l l}
\hline
\textbf{Hyperparameter} & \textbf{Options} \\
\hline
Learning rate    & 0.01, 0.001, 0.0001, 0.00001 \\
Loss function    & MSE, MAE \\
Regularization   & L1, L2, None \\
Batch size       & 16, 32, 64, 128 \\
\hline
\end{tabular}
\end{table}

\pagebreak

\section{Supplementary Notes}

\subsection{Extended Fiala model limitations}
Fiala-UTCI does not account for individual features such as health conditions, age, body weight, activity level, or acclimatization status \cite{Fiala1999} nor for cultural differences in clothing practices and population-level adaptation to the local climatology, meaning that local adjustments to UTCI have been proposed \cite{Kruger2021_Brazil,Ge2017,Park2014}. The Fiala-UTCI model calculates physiological heat transfer for an average individual, assuming them to be 73.4 kg and have a body fat content of 14\%  \cite{Fiala2012}, which is likely to be a substantially different physiology than those most at risk of heat-related harm -- the elderly, children, infants, those with a high BMI, and underlying health conditions.  

Another major limitation is that the Fiala-UTCI model does not account for long-term lag effects of thermal conditions on heat stress. In fact, the factor explaining the highest proportion of thermal sensation differences between the model prediction and survey observations in two cities in Brazil was historical weather information from the three previous days \cite{Brode2021}. 

Furthermore, the model may be inaccurate outside the intended input range of the Fiala-UTCI, which is the same as the training range for Neural-UTCI \cite{Brode2012}. This limitation is shared by Polynomial-UTCI, and to mitigate this effect input parameters are capped to the training range \cite{Broede2021} for both approximating models. Both approximators perform worst at high wind speeds and high relative humidity due to the sparser training grid sampling in these input ranges (Supplementary Figure S2). 

\subsection{Extended discussion of the Grid data}
The 'Grid data' was first published as part of the 2011 operational procedure paper. It is 104,643 samples of synthetic meteorological conditions and physiological  Fiala-UTCI outputs. The exact sampling procedure for building this grid is not described by the paper. However we know \textit{Ta} has 101 samples along the range -50 to 50 C, with values above 0~\textdegree{}C being sampled twice as often. '\textit{Tr - Ta}' has 21 unique values sampled in intervals of 5\textdegree{}C between -30 and 70\textdegree{}C, sampled with an equal number of instances along its range. \textit{va} is sampled at 11 values between 0.5 and 30.3 m/s. The sampling is uneven with variable distances between the values. The values sampled follow a geometric sequence with a ratio of 3/2 (the next term in the sequence is 1.5 times the previous one). Each of the 11 values to be sampled have an equal number of instances. \textit{rH} is sampled unevenly along its range (5-100\%) without a clear pattern. The number of instances for each grid entry is also uneven with about every 20th percentage point represented at a higher frequency than the rest of the samples.  The 5th, 50th, and 100th \textit{rH} percentage points are sampled with an even greater frequency. The grid sampling strategy is not described anywhere, so it is unclear why this strategy was adopted. It is likely that the grid sampling strategy employed by Bröde et al. (2012) \cite{Brode2012} is at least to a large extent the cause of the lower approximator performance of both Neural-UTCI and Polynomial-UTCI at low air temperatures and high wind speeds as compared to higher temperatures and lower wind speeds which are sampled more frequently in the grid data. Additionally, the heteroscedastic nature of the rH error can at least be partially explained by the design of the training data grid. 

Furthermore, having a synthetic grid-type dataset for training poses a challenge for models as the correlations between the input variables are incorrectly represented. For instance, there is no correlation between wind speed and any of the other inputs in the grid data: naturally, as it is synthetically sampled. On the other hand, a random sample from an Earth systems model (ECHAM4 data) reveals a negative correlation between wind speed and vapor pressure, mean radiant temperature, and Ta on the order of -0.4. Because of this lack of information in the training data, when we apply any emulator model to realistic conditions (a distributional shift from the grid data), it is difficult for models to perform well.

\putbib[references]

\end{bibunit}

\end{document}